\documentclass[12pt,draftclsnofoot,onecolumn]{IEEEtran}
\usepackage{graphicx}
\usepackage{amsmath}
\usepackage{amssymb}
\usepackage{subfig}
\usepackage[noadjust]{cite}
\usepackage{float}
\usepackage{algorithm}
\usepackage{algpseudocode}
\usepackage{xcolor}

\DeclareMathOperator*{\argmax}{argmax}
\DeclareMathOperator*{\Real}{Re}
\DeclareMathOperator*{\Imag}{Im}
\DeclareMathOperator*{\sgn}{sign}
\DeclareMathOperator*{\cov}{cov}
\DeclareMathOperator*{\Q}{Q}
\DeclareMathOperator{\spn}{span}
\newcommand{\dd}{\mathop{}\!\mathrm{d}}
\newcommand{\sfrac}[3][0pt]{\genfrac{}{}{}{}{\raisebox{-#1}{$#2$}}{\raisebox{+#1}{$#3$}}}

\newcommand{\gl}[2]{%
  \mathrel{\mathop\gtrless\limits^{#1}_{#2}}%
}

\newtheorem{theorem}{Theorem}
\newtheorem{lemma}{Lemma}

\algnewcommand{\Initialize}[1]{%
  \State \textbf{initialize}
  \Statex \hspace*{\algorithmicindent}\parbox[t]{\linewidth}{\raggedright #1}
}

\begin{document}

\title{A MAP-Based Layered Detection Algorithm and Outage Analysis over MIMO Channels}

\author{Yavuz~Yap{\i}c{\i},~\.{I}smail~G\"{u}ven\c{c},~and~Yuichi~Kakishima
\thanks{Y.~Yap{\i}c{\i}, and \.{I}.~G\"{u}ven\c{c} are with the Department of Electrical and Computer Engineering, North Carolina State University, Raleigh, NC (e-mail:~\{yyapici,iguvenc\}@ncsu.edu).

Y.~Kakishima is with DOCOMO Innovations Inc., Palo Alto, CA 94304 USA (e-mail:~kakishima@docomoinnovations.com)}}%

\maketitle

\begin{abstract}
Efficient symbol detection algorithms carry critical importance for achieving the spatial multiplexing gains promised by multi-input multi-output (MIMO) systems. In this paper, we consider a maximum a posteriori probability (MAP) based symbol detection algorithm, called M-BLAST, over uncoded quasi-static MIMO channels.  Relying on the successive interference cancellation (SIC) receiver, M-BLAST algorithm offers a superior error performance over its predecessor V-BLAST with a signal-to-noise ratio (SNR) gain of as large as $2$~dB under various settings of recent interest. Performance analysis of the M-BLAST algorithm is very complicated since the proposed detection order depends on the decision errors dynamically, which makes an already complex analysis of the conventional ordered SIC receivers even more difficult. To this end, a rigorous analytical framework is proposed to analyze the outage behavior of the M-BLAST algorithm over binary complex alphabets and two transmitting antennas, which has a potential to be generalized to multiple transmitting antennas and multidimensional constellation sets. The numerical results show a very good match between the analytical and simulation data under various SNR values and modulation alphabets.
\end{abstract}


\section{Introduction}\label{sec:intro}
The use of multiple antennas at both the transmitter and the receiver has long been known to greatly increase the achievable rates through spatial multiplexing~\cite{Foschini98OnLim,Telatar99CapMIMO}. Therefore, multi-input multi-output (MIMO) communication has become a key technique in many high data-rate communication technologies including 5G millimeter-wave (mmWave) communications~\cite{Heath14FiveDis,Andrews14What5G,Heath14BeamMul}. Although the conventional purpose for the use of multiple antennas is to combat adverse effects of the channel fading by increasing \textit{diversity}, inherent virtual spatial channels can also be employed as a means for \textit{multiplexing} the independent data streams~\cite{Tse2003DMT}. Bell Labs Layered Space-Time (BLAST) is one of the most famous transmitter architecture to provide spatial multiplexing over MIMO channels~\cite{Foschini96LaySpaTim}. In order to fully exploit the advantage that the vertical BLAST (V-BLAST) technique promises, powerful symbol detection algorithms should be employed at the receiver side, which comes in general with the price of high computational complexity.  

Although the optimal symbol detection over MIMO channels under Gaussian noise perturbation involves maximum likelihood (ML) rule, the resulting complexity grows exponentially with the increasing number of antennas and size of the modulation alphabet~\cite{Hanzo2015MIMODet,Larsson2009MIMODet,Schober2009MIMOLitSur}. In~\cite{Viterbo99UniLatCod}, the sphere decoding (SD) algorithm is introduced with a promise of near-ML error performance at a relatively low complexity, which is unfortunately not polynomial-time under any settings of interest~\cite{Hassibi05OnSphereDecI,Hassibi04ModSphereDec}. Towards reducing the decoding complexity of the SD algorithm, several extensions are proposed with various strategies including reordering the channel matrix and searching over a subset of the complete constellation~\cite{Barbero2008}, decoding the real and imaginary parts independently~\cite{Ayanoglu2009RedComSph}, and introducing an interference based cost function~\cite{Romano2013}.

The original V-BLAST detection algorithm relies on the successive interference cancellation (SIC), where symbols are detected sequentially based on their a priori average signal-to-noise ratio (SNR)~\cite{Foschini99VBLAST}. Because the SIC is highly prone to the error propagation, the symbol detection order and its implementation have been received much attention in the literature. In~\cite{Wubben2003}, QR factorization based detection order is adopted whereas \cite{Liu2008,Xia2009} modifies the well known \textit{fast recursive algorithm} (FRA)~\cite{Benesty2003FRA} to reduce the computationally complexity of the original V-BLAST. In~\cite{Xu2005,Kim2006,Lee2007,Lee2009}, log-likelihood ratio (LLR) based approaches are considered in computing the symbol detection order without any performance analysis, which are extensively used in soft-decision based iterative decoders for V-BLAST receiver~\cite{Yuk2003,Lee2006,Singer2010}. Recently, the dependency of the optimal detection order on the power and rate allocation strategies is considered in~\cite{Loyka2014OptDetOr}, and the ordered SIC (OSIC) based low-complexity receivers are studied in~\cite{Medra2016,Mashed2015}.

This paper, which is substantially enhanced version of~\cite{Yapici2017OutAnaMBLAST}, considers an improvement over the conventional V-BLAST detection algorithm, and provide a rigorous analytical performance evaluation for the proposed SIC receiver. This new algorithm, which is considered previously by~\cite{YapiciMSc,Yapici2007AnImp}, offers a novel symbol detection order for the SIC receiver inspired by the maximum a posteriori probability (MAP) rule and is therefore referred to as MAP-based BLAST (M-BLAST). In contrast to the original V-BLAST algorithm, which determines the detection order statically based on the channel matrix only, the new M-BLAST algorithm orders the subchannels dynamically by taking into account the resulting symbol decision error, as well. The key contributions can be summarized as follows: 
\begin{itemize}
\item[i.] The M-BLAST algorithm is shown to have a superior performance over the original V-BLAST under various settings of recent interest including arbitrary size of antenna arrays and modulation alphabets, with as large as $2$~dB SNR gain in some certain settings. Considering that our simulation settings involve point-to-point Rician fading together with massive and correlated MIMO channels, M-BLAST can have promising potential for applications in next-generation wireless systems, such as the massive MIMO and correlated MIMO in 5G mmWave networks.
\item[ii.] We rigorously derive key metrics for the MAP receiver which are employed to determine the proposed SIC detection order. In particular, the subchannel SNR expressions and the distribution of the modified noise are derived rigorously with direct relation to the channel matrix, which are used as the two major definitions in the analytical performance evaluation, as well. We also show that the proposed detection order comes with an additional complexity which is linear with the modulation alphabet size $|\mathcal{A}|$.
\item[iii.] We provide a rigorous performance analysis for the M-BLAST algorithm, which has not been studied in the literature before. For this analysis, we consider two transmit and arbitrary number of receive antennas, as in~\cite{Loyka2004PerAnVB}. In addition, because the proposed detection order is a nonlinear function of the channel matrix, we consider generalized complex binary alphabets and derive an equivalent linear expression for the detection order. The distributions of the decision-dependent random variables introduced by this equivalent ordering rule are characterized rigorously under both the perfect and imperfect detection scenarios. The outage probabilities at each detection layer of the SIC receiver are then derived based on this new framework, which are shown to follow the simulation data successfully. We note that this derivation can be extended to the arbitrary number of transmit antennas and multidimensional constellations by following the strategy of~\cite{Loyka2008OutErr}, as a future work.
\end{itemize}

The rest of the paper is organized as follows. In Section~\ref{sec:system_model}, system model is introduced. The algorithm flow of the M-BLAST is overviewed in Section~\ref{sec:mblast_algorithm}, and an equivalent rule for the proposed detection order is presented in Section~\ref{sec:mblast_binary_ordering_rule}. The distributions of the decision-dependent random variables are derived in Section~\ref{sec:statistical_analysis}. The analytical outage expressions are derived in Section~\ref{sec:outage_analysis} with the numerical results in Section~\ref{sec:numerical_results}. The paper concludes with Section~\ref{sec:conclusion}.

\textit{Notations:} $(\cdot)^H$, $(\cdot)^*$, $(\cdot)^+$, $\|\cdot\|$, $|\cdot|$, $\mathbb{E}\{\cdot\}$, $\cov(\cdot)$ denote Hermitian transpose, complex conjugation, Moore-Penrose pseudoinverse, Euclidean norm, absolute-value norm, statistical expectation, and covariance operators, respectively. The symbol $\rightarrow $ is used to denote the limiting behaviour. $(\textbf{A})_i$ and $[\textbf{A}]_{nn}$ are the $i$th row and $n$th diagonal element of matrix $\textbf{A}$, respectively. $\textbf{I}_M$ is the $M{\times}M$ identity matrix, and $\delta (a,b)$ is Kronecker delta taking $1$ if $a\,{=}\,b$, and $0$ otherwise. $\mathcal{N}\left(\mu,\sigma^2\right)$ and $\mathcal{CN}\left(\mu,\sigma^2\right)$ denote the real and complex valued Gaussian distribution, respectively, with the mean $\mu$ and the variance $\sigma^2$.

\section{System Model}\label{sec:system_model}
We consider a single-user communication system with $M$ transmitter and $N$ receiver antennas over a frequency-flat spatially uncorrelated Rayleigh fading channel. The discrete-time complex baseband equivalent channel model is given by
\begin{equation}\label{eqn:mimo_channel_model_1}
   \textbf{r} = \textbf{H} \, \textbf{x} + \textbf{v} \, ,
\end{equation}
where $\textbf{r} \in \mathbb{C}^{N \times 1}$ is the received vector of complex channel observations, and $\textbf{H} \in \mathbb{C}^{N \times M}$ is an $N{\times}M$ channel matrix with independent and identically distributed (iid) complex Gaussian entries with zero mean and $\sigma_h^2$ variance, denoted as $h_{nm}\,{\sim}\,\mathcal{CN}(0,\sigma_h^2)$. The transmitted data vector $\textbf{x} \in \mathbb{C}^{M \times 1}$ is composed of iid symbols chosen from a modulation alphabet $\mathcal{A}$ of the cardinality $|\mathcal{A}|$. The additive white noise $\textbf{v} \in \mathbb{C}^{N \times 1}$ has circularly symmetric complex Gaussian entries with zero mean and variance $\sigma_v^2$, denoted jointly as $\textbf{v}\,{\sim}\,\mathcal{CN}(\textbf{0},\sigma_v^2 \textbf{I}_{N})$. The channel matrix $\textbf{H}$ and the noise variance $\sigma_v^2$ are assumed to be known perfectly at the receiver.

Based on (\ref{eqn:mimo_channel_model_1}), the received signal at any of the receiving antennas can be given as
\begin{equation}\label{eqn:mimo_channel_model_2}
    r_{n} = \displaystyle\sum_{m = 1}^{M} h_{nm} \, x_m + v_n, \quad n = 1,2,\dots,N,
\end{equation}
and, the average received SNR at any of the receiving antenna is accordingly given as
\begin{equation}\label{eqn:received_snr}
    \gamma  = \frac{M \, \mathbb{E}\{|h_{nm}|^2\} \, \mathbb{E}\{|x_{m}|^2\}} {\mathbb{E}\{|v_{n}|^2\}} = \frac{E_\textrm{s}}{\sigma_v^2} M \sigma_h^2 \, ,
\end{equation}
where $E_\textrm{s} = \mathbb{E}\{|x_{m}|^2\}$ is the transmitted symbol energy.

\section{M-BLAST Symbol Detection Algorithm}\label{sec:mblast_algorithm}
In this section, we review the M-BLAST algorithm with the discussion of the associated complexity. We also derive the exact expression of the conditional a posteriori probability and identify the distribution of the modified noise to characterize the post-processing SNR.
\subsection{Algorithm}\label{sec:mblast_defn}
Let $(k_1,k_2,\dots,k_{M})$ be a permutation of the index set $(1,2,\dots,M)$. In this permuted set, each entry $k_i$ stands for the index of the transmitted symbol detected at the $i$th layer, and is determined by the Algorithm~\ref{alg:mblast}, as follows.
\begin{algorithm}
\caption{M-BLAST Algorithm}\label{alg:mblast}
\begin{algorithmic}[1]
\State{\textbf{Initialize:}}
\State\hspace{\algorithmicindent $\textbf{W}_{1} \gets \textbf{H}^+$} \label{line:mblast_1}
\State\hspace{\algorithmicindent $\textbf{r}_{1} \gets \textbf{r}$} \label{line:mblast_2}
\For{$i \gets 1 $ \textbf{to} $M$}
\State $\textbf{y}_{i} \gets \textbf{W}_{i} \, \textbf{r}_{i}$ \label{line:mblast_3}
\State $\textbf{s}_{i} \gets \Q \left(\textbf{y}_{i} \right)$ \label{line:mblast_4}
\State $k_{i} \gets \argmax_{ j \, \notin \, \mathcal{I}_{i{-}1}} \left\lbrace p_{ij} \right\rbrace$ \label{line:mblast_5}
\State $\hat{x}_{k_{i}} \gets s_{i\,k_i}$ \label{line:mblast_6}
\State $\textbf{r}_{i{+}1} \gets \textbf{r}_{i} - \hat{x}_{k_{i}} \textbf{h}_{k_{i}}$ \label{line:mblast_7}
\State $\textbf{W}_{i{+}1} \gets \left( \textbf{H}_{\overline{k}_{i}} \right)^+$ \label{line:mblast_8}
\EndFor
\end{algorithmic}
\end{algorithm}
  
In the algorithm flow, $\Q(\cdot)$ denotes a quantizer mapping each element of its argument vector to the nearest constellation point in the alphabet $\mathcal{A}$, the set $\mathcal{I}_{i{-}1} {=} \left\lbrace k_1,\dots,k_{i{-}1} \right\rbrace$ includes the set of indices processed by the algorithm before the $i$th layer, $\textbf{H}_{\overline{k}_{i}}$ is the modified channel matrix obtained by setting the columns $\left\lbrace k_1,...,k_{i} \right\rbrace$ to zero, $\textbf{h}_{k_{i}}$ is the $k_i$th column of the channel matrix. This algorithm essentially differentiates from the original V-BLAST in the way the detection order is evaluated, which is given by (\algref{alg:mblast}{line:mblast_6}) for the M-BLAST and defined as $k_{i}\,{=}\,\argmax_{ j \, \notin \, \mathcal{I}_{i{-}1}} \| \left( \textbf{W}_i \right)_j \|^2$ for the V-BLAST~\cite{Foschini99VBLAST}.

At the $i$th layer, the M-BLAST algorithm nulls out the interference from yet to be detected symbols as in (\algref{alg:mblast}{line:mblast_3}), and obtains the output vector $\textbf{y}_i$ which corresponds to observations in each of $M$ virtual subchannels. The symbol-wise demodulation is then applied to obtain the tentative decisions $s_{ij}\,{=}\,\Q\left(y_{ij}\right)$, which is shown jointly in (\algref{alg:mblast}{line:mblast_4}). Note that, the nulling and demodulation steps of the V-BLAST algorithm are performed only for the subchannel chosen to be detected. Based on these tentative decisions, the \textit{reliability measures} are given as
\begin{align}\label{eqn:reliability_measures}
p_{ij} &= \frac{f_{ij} \left( y_{ij} \, | \, s_{ij} \right) }{\displaystyle\sum_{a \in \mathcal{A}} f_{ij} \left( y_{ij} \, | \, a \right)} \, ,
\end{align}
where $f_{ij}$ is the complex Gaussian density function given by
\begin{align}\label{eqn:likelihood_fnc}
	f_{ij}\left(y_{ij} \, | \, x_j \right) &= \frac{1}{\pi\sigma_{ij}^2} \exp \left\lbrace -\frac{1}{\sigma_{ij}^2} \left\|y_{ij}-x_j \right\|^2 \right\rbrace \, ,
\end{align}
with the post-processing noise variance
\begin{align}\label{eqn:modf_noise_var}
\sigma_{ij}^2 = \sigma_v^2\| (\textbf{W}_i)_j \|^2 \, .
\end{align} 

The algorithm chooses the subchannel with the largest $p_{ij}$ as in (\algref{alg:mblast}{line:mblast_5}) among those which have not been processed yet. The symbol decision is then obtained by selecting the $k_i$th element of the tentative decision vector, denoted as $s_{i\,k_i}$ in (\algref{alg:mblast}{line:mblast_6}). The rest of the algorithm, which is common with V-BLAST~\cite{Foschini99VBLAST}, involves the symbol cancellation to produce the modified received vector $\textbf{r}_{i+1}$, and the update of the nulling matrix steps, shown in (\algref{alg:mblast}{line:mblast_7}) and (\algref{alg:mblast}{line:mblast_8}), respectively.

\begin{figure}[!h]
\centering
\includegraphics[width=0.7\textwidth]{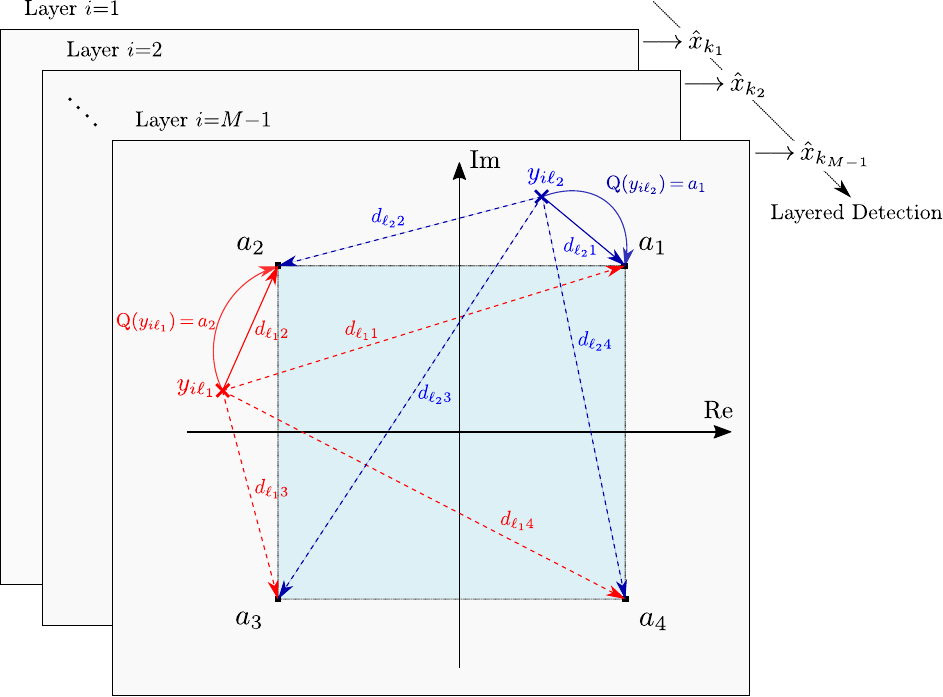}
\caption{A sketch for the detection order of the M-BLAST at the $i\,{=}\,(M{-}1)$th layer with $\mathcal{A}\,{=}\,\{a_m\}_{m\,{=}\,1}^4$. The indices of the undetected symbols are $\ell_1$ and $\ell_2$. The Euclidean distances $d_{\ell_jm}\,{=}\,\|y_{i\ell_j}{-}a_m\|$ are shown with dotted lines except for the shortest quantization path, $\Q(y_{i\ell_j})$.}
\label{fig:mblast_order}
\end{figure}

In Fig.~\ref{fig:mblast_order}, a simplified sketch is given for the processing of the subchannel order of the M-BLAST at the $i\,{=}\,(M{-}1)$th layer. The reliability measures $p_{i\ell_1}$ and $p_{i\ell_2}$ are computed for the unprocessed subchannels $\ell_1$ and $\ell_2$ according to (\ref{eqn:reliability_measures}), and compared as in (\algref{alg:mblast}{line:mblast_5}) to choose the subchannel to be detected at this layer. Note that, the computation of the reliability measures involves the Euclidean norms denoted as $d_{\ell_jm}\,{=}\,\|y_{i\ell_j}{-}a_m\|$ as well as the modified noise $\sigma_{i\ell_j}^2$ given in (\ref{eqn:modf_noise_var}), in contrast to V-BLAST which employs $\sigma_{i\ell_j}^2$ only.

\subsection{Conditional A Posteriori Probability and Probability Measures}
The detection order of the M-BLAST is based on a comparison of conditional \textit{a posteriori} probabilities, $\Pr\left( x_j\,{=}\,s_{ij}\,{|}\,y_{ij} \right)$, of tentative decisions for each undetected symbols, and selecting the one with the largest of these probabilities, in a similar way to the MAP rule. 

The conditional a posteriori probability of $s_{ij}$ being the transmitted symbol in subchannel $j$ given that $y_{ij}$ is observed at the output of that subchannel is given as
\begin{align}\label{eqn:aposteriori_prob}
	\Pr\left( x_j\,{=}\,s_{ij}\,{|}\,y_{ij} \right) &= \frac{f_{ij} \left( y_{ij}\,{|}\,x_j\,{=}\,s_{ij} \right) \Pr\left( x_j\,{=}\,s_{ij} \right)}{\displaystyle\sum_{a\in \mathcal{A}} f_{ij} \left(y_{ij}\,{|}\,x_j\,{=}\,a \right)\Pr\left( x_j\,{=}\,a \right)} \,,
\end{align}
where $f_{ij}$ is the complex Gaussian probability distribution function (pdf) given in (\ref{eqn:likelihood_fnc}). In order to verify this pdf and the associated noise variance in (\ref{eqn:modf_noise_var}), consider the nulling step of M-BLAST by employing the system model in (\ref{eqn:mimo_channel_model_1}) and the accumulated expression of $\textbf{r}_i$, as follows
\begin{align}
\textbf{y}_{i} &= \textbf{W}_{i} \!\! \displaystyle\sum_{\substack{j = 1 \\ j \notin \mathcal{I}_{i{-}1}}}^{M} \textbf{h}_{j} x_{j}  + \textbf{W}_{i}  \displaystyle\sum_{j = 1}^{i{-}1} \, \textbf{h}_{k_j} \! \left(x_{k_j}{-}\hat{x}_{k_j}\right) + \textbf{W}_{i} \,\textbf{v} . \label{eqn:mblast_nulling_2}
\end{align}
Employing the modified channel matrix $\textbf{H}_{\overline{k}_{i-1}}$ gives 
\begin{align}
   \textbf{y}_{i} &= \textbf{W}_{i} \, \textbf{H}_{\overline{k}_{i{-}1}} \textbf{x}  + \textbf{W}_{i} \, \displaystyle\sum_{j=1}^{i{-}1} \, \textbf{h}_{k_j} \Delta x_{k_j} + \tilde{\textbf{v}}_i \,, \label{eqn:mblast_nulling_3}
\end{align}
where $\Delta x_{k_j} {=} \, x_{k_j} {-} \hat{x}_{k_j}$ is the decision error made at the $k_j$th layer, and $\tilde{\textbf{v}}_i = \textbf{W}_{i} \,\textbf{v}$ is the modified noise vector at the $i$th layer with the covariance matrix
\begin{align}\label{eqn:noise_cov}
\textbf{C}_{\tilde{\textbf{v}}_i} = \mathbb{E}\{\tilde{\textbf{v}}_i\tilde{\textbf{v}}_i^H\} = \sigma_v^2 \, \textbf{W}_{i} \, \textbf{W}_{i}^H .
\end{align}

If the channel matrix $\textbf{H}$ has full-column rank as in rich-scattering channels, we have $\textbf{W}_{i} \, \textbf{H}_{\overline{k}_{i{-}1}} {=} \, \textbf{I}_M$, which directly follows from (\algref{alg:mblast}{line:mblast_8}). Assuming perfect detection for the previous layers, i.e., $\Delta x_{k_j} {=} \, 0$ for $j {=} 1,\dots,i{-}1$, (\ref{eqn:mblast_nulling_3}) represents a set of $M{-}i{+}1$ subchannels given as
\begin{align}
y_{ij} &= x_j + \tilde{v}_{ij} \,, \label{eqn:subchannel_defn}
\end{align}
for $j \notin \mathcal{I}_{i-1} {=} \left\lbrace k_1,\dots,k_{i-1} \right\rbrace$, which verifies the density in (\ref{eqn:likelihood_fnc}) for a given $x_j$ and channel matrix. 

{Finally, we readily obtain the reliability measures in (\ref{eqn:reliability_measures}) since the symbols in (\ref{eqn:aposteriori_prob}) are equally likely. The overall derivation shows that the M-BLAST takes into account decision errors while computing the detection order, whereas the V-BLAST considers only the post-processing noise variance, which ends up with the superior error performance of the M-BLAST. Note that a posteriori probability in (\ref{eqn:aposteriori_prob}) is conditional since the perfect detection is assumed for the previous layers, which can be satisfied for sufficiently large SNR.}

\subsection{Complexity Comparison}\label{sec:complexity}
In this section, we compare the complexity of the M-BLAST and V-BLAST algorithms in the order of magnitudes representation via $\mathcal{O(\cdot)}$ notation, since different implementations of these algorithms may affect the exact number of floating point operations~\cite{Hunger2007FloPoi}. 

As discussed in Section~\ref{sec:mblast_defn}, the difference between the M-BLAST and the V-BLAST algorithms appears while running the steps of the detection order computation and the nulling. At the $i$th layer, both algorithms compute the post-processing noise variances given by~\eqref{eqn:modf_noise_var} for $M{-}i{+}1$ unprocessed subchannels, which can be considered as $\mathcal{O}(M N)$ computations. While the V-BLAST is using these values directly to determine the detection order, the M-BLAST employs them in the computation of the reliability measures in~\eqref{eqn:reliability_measures} for all $M{-}i{+}1$ unprocessed subchannels, which brings an additional complexity of $\mathcal{O}(M |\mathcal{A}|)$. 

On the other hand, the nulling of interference from the unprocessed subchannels requires $\mathcal{O}(N)$ computations for the V-BLAST whereas the respective complexity for the M-BLAST is $\mathcal{O}(MN)$ since it considers all the unprocessed subchannels at each layer. As a result, the M-BLAST requires $\mathcal{O}(M |\mathcal{A}|)$ more computations as compared to the V-BLAST while calculating the detection order, and $\mathcal{O}(MN)$ computations for nulling (which is $\mathcal{O}(N)$ for the V-BLAST). In addition, both algorithms involve the common symbol cancellation and nulling steps with the complexity $\mathcal{O}(N)$ and $\mathcal{O}(M^2\max\{M,N\})$, respectively, at each detection layer~\cite{Hunger2007FloPoi}.

\subsection{Post-Processing SNR}\label{sec:post_processing_snr}
When the channel matrix $\textbf{H}$ has full-column rank, the nulling matrix at the $1$st layer becomes $\textbf{W}_{1} = \left(\textbf{H}^H \textbf{H} \right)^{-1} \textbf{H}^H$. The diagonal elements of the covariance matrix for the modified noise vector given in (\ref{eqn:noise_cov}) can then be given as
\begin{align}
\sigma_{1j}^2  &= \sigma_v^2 \, \left[ \left( \textbf{H}^H \textbf{H} \right)^{-1} \right]_{jj} = \frac{\sigma_v^2}{\left\| \textbf{h}_j^{\bot} \right\|^2} \,,\label{eqn:noise_var}
\end{align}
for $j = 1,2,\dots,M$ \cite{Varanasi2008SpMuI}. The notation $\textbf{h}_j^{\bot}$ denotes the component of $\textbf{h}_j$ in the null space of the matrix $\textbf{H}_j$ which is obtained from $\textbf{H}$ by simply striking the $j$th column out. We therefore have
\begin{align}
\textbf{h}_j^{\bot} &= \textbf{P}_{\textbf{H}_j}^{\bot} \textbf{h}_j \,,\label{eqn:orth_column}
\end{align} 
where $\textbf{P}_{\textbf{H}_j}^{\bot} = \textbf{I} - \textbf{H}_j \left( \textbf{H}_j^H \textbf{H}_j\right)^{-1} \textbf{H}_j^H$ is the orthogonal projection matrix onto the null space of $\textbf{H}_j$. The effective SNR in each subchannel of (\ref{eqn:subchannel_defn}) is then given for the $1$st layer as follows
\begin{align}\label{eqn:subchannel_snr}
\gamma_s &= \frac{E_\textrm{s}}{\sigma_{1j}^2} = \frac{E_\textrm{s}}{\sigma_{v}^2} \, \left\| \textbf{h}_j^{\bot} \right\|^2 .
\end{align}
This post-processing SNR definition implies that $\textbf{h}_j^{\bot}$ may also be interpreted as a random vector whose square-norm is proportional to the signal energy after nulling the interference from yet to be detected symbols out, as argued in~\cite{Loyka2004PerAnVB,Loyka2008OutErr}. 

Note that, although $\tilde{v}_{ij}$ is complex Gaussian for a given channel realization $\textbf{H}$, it is no more exactly Gaussian when the random nature of $\textbf{H}$ is incorporated into the analysis. However, because each element of the modified noise vector is a weighted sum of uncorrelated noise samples, i.e., $\tilde{v}_{ij} = \left( \textbf{W}_i \right)_j \textbf{v}$, the random variable $\tilde{v}_{ij}$ can be safely approximated as Gaussian as the number of receiving antennas $N$ increases, from the well-known Central Limit Theorem~\cite{PapoulisProb}. 

Because channel gains have equal variance of $\sigma_h^2$, $\|\textbf{h}_j^{\bot}\|^2$ has chi-square distribution with $2(N{-}M{+}i)$ degrees of freedom at the $i$th layer, i.e., $\|\textbf{h}_j^{\bot}\|^2 \, {\sim} \, \chi_{2\left(N-M+i\right)}^2$~\cite{Loyka2004PerAnVB}. Employing the average value $\mathbb{E} \left\lbrace \|\textbf{h}_j^{\bot}\|^2 \right\rbrace$ instead of the instantaneous value $\|\textbf{h}_j^{\bot}\|^2$ in (\ref{eqn:noise_var}), the noise variance $\sigma_{1j}^2$ can be approximated with a relation to the input SNR given in (\ref{eqn:received_snr}) as follows
\begin{align}
\bar{\sigma}^2  &= \frac{\sigma_v^2}{\left(N{-}M{+}1\right) \sigma_h^2} = \frac{M}{N{-}M{+}1} \frac{E_\textrm{s}}{\gamma} \,. \label{eqn:noise_var_ave_2}
\end{align}
As a result, the distribution of the modified noise can be well approximated by $\mathcal{CN}\left(0,\bar{\sigma}^2 \right)$.   
 
\section{M-BLAST Ordering Rule over Binary Alphabets}\label{sec:mblast_binary_ordering_rule}
When transmitted symbols are chosen from a general binary alphabet, the reliability measures and detection order of the M-BLAST can be reformulated in a more compact form. 

\begin{theorem}\label{theorem:binary_ordering_rule}
Assuming a binary modulation alphabet $\mathcal{A}\,{=}\,\{a_1,a_2\}$  with $a_1,a_2 \in \mathbb{C}$, the reliability measure for the $j$th subchannel is given as
\begin{align}\label{eqn:pij_binary}
p_{ij} &= \left[ 1 + \exp \left\lbrace {-}\frac{2 \Real\left\lbrace \, y_{ij} \, \Delta a^* \, \right\rbrace}{\sigma_{ij}^2} \Delta \delta \left( s_{ij} \right) \right\rbrace \right]^{-1} \, ,
\end{align}
and, the decision rule for the ordering of subchannels when $M\,{=}\,2$ is formulated as
\begin{align}\label{eqn:binary_ordering}
\left\| \textbf{h}_1 \right\|^2 u_1 \gl{E_1}{E_2} \; \left\| \textbf{h}_2 \right\|^2 u_2 \, .
\end{align} 
In the ordering rule of (\ref{eqn:binary_ordering}), $u_j$ is the decision-dependent random variable given as
\begin{align}\label{eqn:uj}
u_j = \Real\left\lbrace y_{1j} \, \Delta a^* \right\rbrace \Delta \delta \left( s_{1j} \right) \,,
\end{align}
where $\Delta a\,{=}\,a_1{-}a_2$ is the difference of constellation points in the alphabet $\mathcal{A}$, $\Delta \delta \left( s_{ij} \right) = \delta(s_{ij},a_1) {-} \delta(s_{ij},a_2)$ is the Kronecker delta difference which takes ${+}1$ or ${-}1$ depending on the assignment being made for $s_{ij}$, and $E_m$ is the subchannel ordering event given as
\begin{equation}\label{eqn:ordering_events}
E_m = \left\{
  \begin{array}{lr}
    \left(k_1,k_2\right) = \left(1,2\right), \;\; \text{if} \;\; m = 1 \\
    \left(k_1,k_2\right) = \left(2,1\right), \;\; \text{if} \;\; m = 2 
  \end{array} \, .
\right.
\end{equation}
\end{theorem}

\begin{IEEEproof}
See Appendix~\ref{app:binary_ordering_rule}.
\end{IEEEproof}

Note that the M-BLAST ordering rule in (\ref{eqn:binary_ordering}) definitely relies on the channel observations $y_{1j}$ through random variables $u_j$, as well as the channel matrix through the column vectors $\textbf{h}_j$. In contrast, V-BLAST ordering rule considers solely the channel matrix, hence $u_j$ can be interpreted as unity in V-BLAST, i.e., $u_j = 1$. Note also that (\ref{eqn:binary_ordering}) is linear in the square-norm of the column vectors $\textbf{h}_j$ and completely defines the subchannel order at the $1$st detection layer. As a final remark, the decision-dependent random variable $u_j$ can be interpreted as a measure for the dependency of the M-BLAST ordering rule to the channel observations.

\section{Statistical Behaviour of $u_j$ and the Ratio $u=u_2/u_1$}\label{sec:statistical_analysis}
In this section, we derive the distribution of the decision-dependent random variables $u_j$ for $j\,{=}\,1,2$, and their ratio $u\,{=}\,u_2/u_1$. This analysis is key to the performance analysis of the M-BLAST in the next section as it employs the decision rule in (\ref{eqn:binary_ordering}) involving these random variables. To this end, we elaborate (\ref{eqn:uj}) in the following for the real-valued binary phase shift-keying (BPSK) and complex-valued binary frequency shift-keying (BFSK) alphabets, separately. 

\begin{lemma}\label{lemma:uj_bpsk}
For BPSK modulation with the alphabet $\mathcal{A}_P\,{=}\,\{a_1,{-}a_1\}$ where $a_1\,{\in}\,\mathbb{R}^{+}$, we have $\Delta a \,{=}\,2a_1$ and $\Delta \delta \left( s_{1j} \right)\,{=}\,\delta(s_{1j},a_1){-}\delta(s_{1j},{-}a_1)$, where it is easy to see that $\Delta \delta \left( s_{1j} \right)\,{=}\,\sgn(s_{1j})$. After cancellation of $\Delta a$ at both side of (\ref{eqn:binary_ordering}), we obtain the following effective expression 
\begin{align}\label{eqn:uj_bpsk}
u_j = \Real\left\lbrace y_{1j} \right\rbrace \sgn \left( s_{1j} \right).
\end{align}
\end{lemma}

\begin{lemma}\label{lemma:uj_bfsk}
For BFSK modulation with the alphabet $\mathcal{A}_F\,{=}\,\{a_1,ja_1\}$ where $a_1\,{\in}\,\mathbb{R}^{+}$, we have $\Delta a\,{=}\,a_1(1{-}j)$ and $\Delta \delta \left( s_{1j} \right)\,{=}\,\delta(s_{1j},a_1){-}\delta(s_{1j},ja_1)$. Using the simplified expression $\Delta \delta \left( s_{1j} \right)=s_{1j}^2 / a_1^2$, which is easy to develop, we have
\begin{align}\label{eqn:uj_bfsk}
u_j = \Real\left\lbrace y_{1j} (1+j) \right\rbrace s_{1j}^2/a_1.
\end{align}
\end{lemma}

\subsection{Statistical Analysis of $u_j$}\label{sec:statistics_uj}
In this section, the statistical behaviour of the random variable $u_j$ will be investigated for the perfect and the imperfect symbol detection cases, in sequence. We begin the analysis by assuming perfect symbol detection where the tentative symbol decisions in each of two subchannels are perfectly detected. The distribution of $u_j$ is then given in the following theorem.  
\begin{theorem}\label{theorem:uj_perfect}
Assuming perfect symbol detection such that $s_{1j}\,{=}\,x_j$ for $j\,{=}\,1,2$, the distributions of $u_j$'s for both BPSK and BFSK modulations are uncorrelated and real-valued Gaussian given as $\mathcal{N}\left(a_1,\bar{\sigma}^2\!/2 \right)$ and $\mathcal{N}\left(a_1^2,a_1^2\bar{\sigma}^2 \right)$, respectively, where $\bar{\sigma}^2$ is defined in~(\ref{eqn:noise_var_ave_2}).
\end{theorem}
\begin{IEEEproof}
See Appendix~\ref{app:distribution_uj_perfect}.
\end{IEEEproof}

In order to investigate the effect of more realistic imperfect symbol detection assumption on the results of Theorem~\ref{theorem:uj_perfect}, we now employ the optimal minimum distance detector~\cite{ProakisDigiComm} to obtain the actually detected symbols, and present the modified distribution as follows. 
\begin{theorem}\label{theorem:uj_imperfect}
Assuming imperfect symbol detection where the optimal minimum distance receiver~\cite{ProakisDigiComm} is employed, the pdf of $u_j$'s for BPSK and BFSK modulations are given as
\begin{align}
f_{u_{j}|\textrm{P}} (x) &= f_{\mathcal{N}}\left(x,a_1,\bar{\sigma}^2\!/2 \right) \left[ 1{+}\exp\left\lbrace {-}\frac{4 a_1}{\bar{\sigma}^2}x \right\rbrace \right] \,,  &\text{(for BPSK)} \label{eqn:pdf_imper_bpsk} \\
f_{u_{j}|\textrm{F}} (x) &= f_{\mathcal{N}}(x,a_1^2,a_1^2\bar{\sigma}^2) \left[ 1{+}\exp\left\lbrace {-}\frac{2x}{\bar{\sigma}^2} \right\rbrace \right] \,,  &\text{(for BFSK)} \label{eqn:pdf_imper_bfsk}  
\end{align}
where $f_{\mathcal{N}}(x,\mu,\sigma^2)$ denotes the pdf of the random variable $x$ distributed by $\mathcal{N}\left(\mu,\sigma^2 \right)$.
\end{theorem}
\begin{IEEEproof}
See Appendix~\ref{app:distribution_uj_imperfect}.
\end{IEEEproof}

We observe that the exponential terms in (\ref{eqn:pdf_imper_bpsk}) and (\ref{eqn:pdf_imper_bfsk}) get smaller to zero as $x$ increases, and they eventually vanish at sufficiently large SNR ($\bar{\sigma}^2\,{\ll}\,1$) for which \eqref{eqn:pdf_imper_bpsk} and \eqref{eqn:pdf_imper_bfsk} become equal to those in Theorem~\ref{theorem:uj_perfect}. We therefore conclude that the distribution of $u_j$ under the realistic imperfect symbol detection assumption approaches or even becomes identical to that of the ideal perfect symbol detection assumption for practical settings. 

\begin{figure}[!h]
\centering
\subfloat[$\gamma = -5$~dB]{\includegraphics[width=0.5\textwidth]{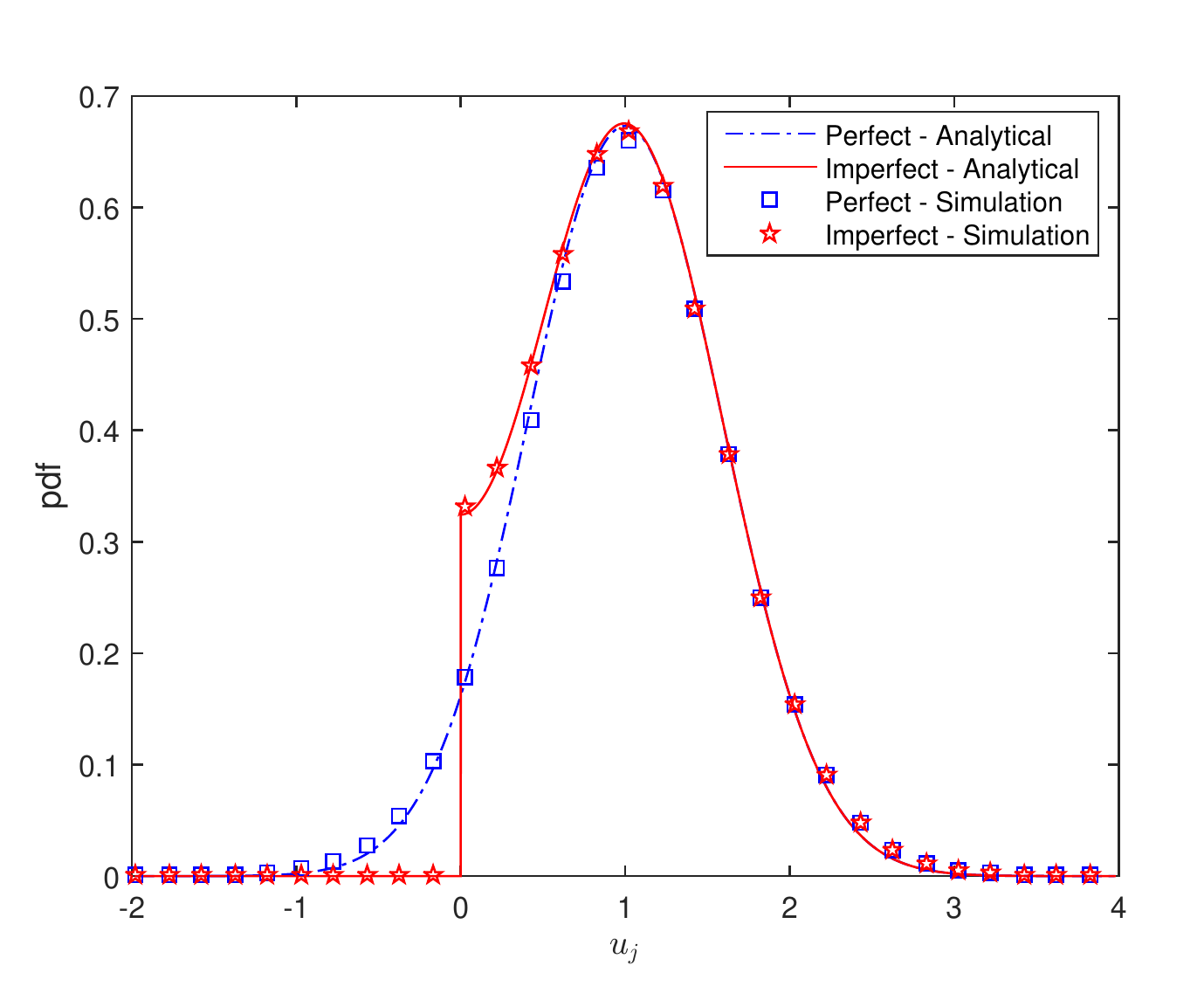}
\label{fig:pdf_u1_n_10_bpsk_-5dB}}
%\hfil
\subfloat[$\gamma = 0$~dB]{\includegraphics[width=0.5\textwidth]{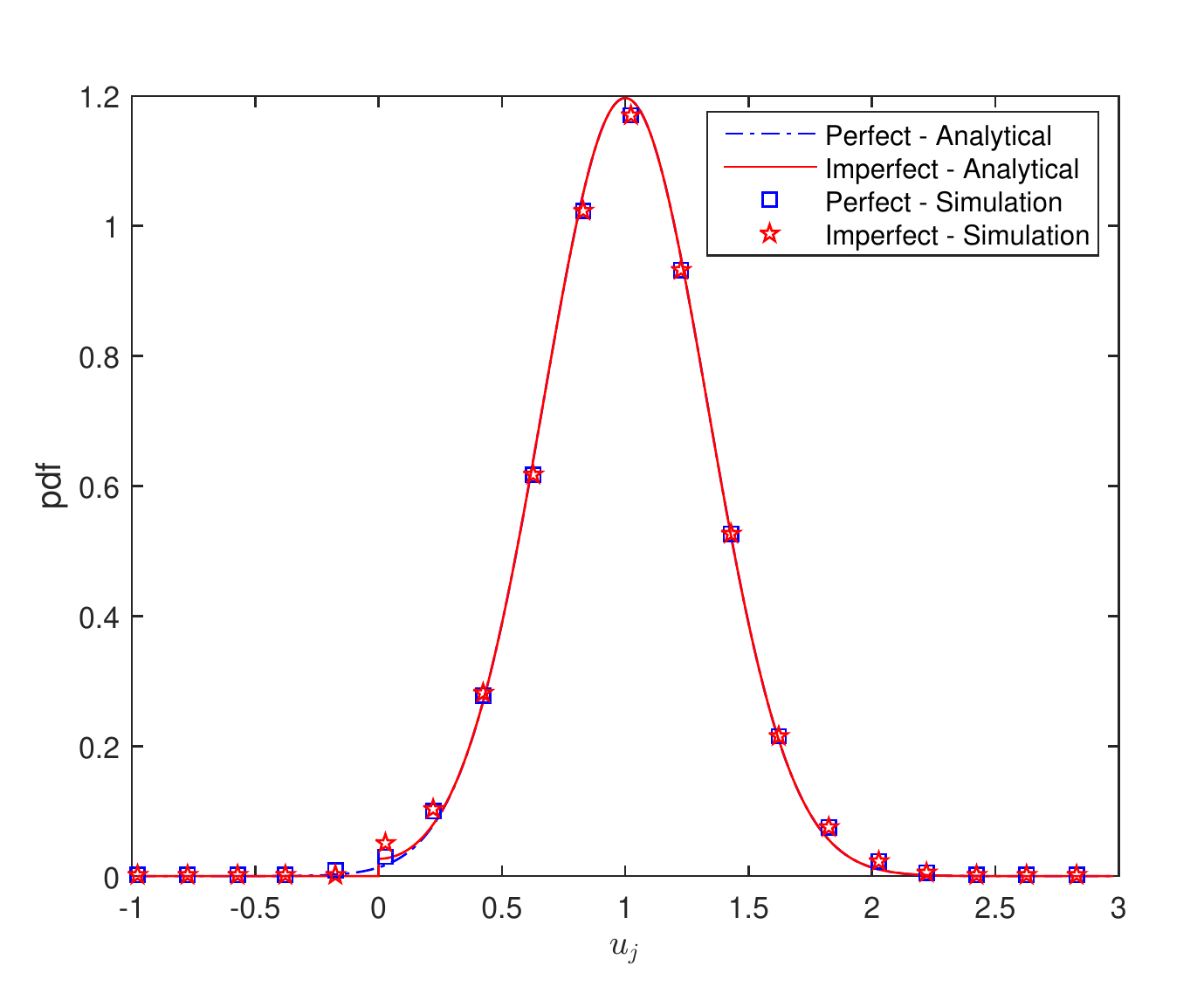}
\label{fig:pdf_u1_n_10_bpsk_0dB}}
\caption{Analytical and simulation results for the pdf of $u_j$ under perfect and imperfect symbol detection cases assuming BPSK modulation and $N_10$.}
\label{fig:pdf_u1_n_10_bpsk}
\end{figure}

\begin{figure}[!h]
\centering
\subfloat[$\gamma = 0$~dB]{\includegraphics[width=0.5\textwidth]{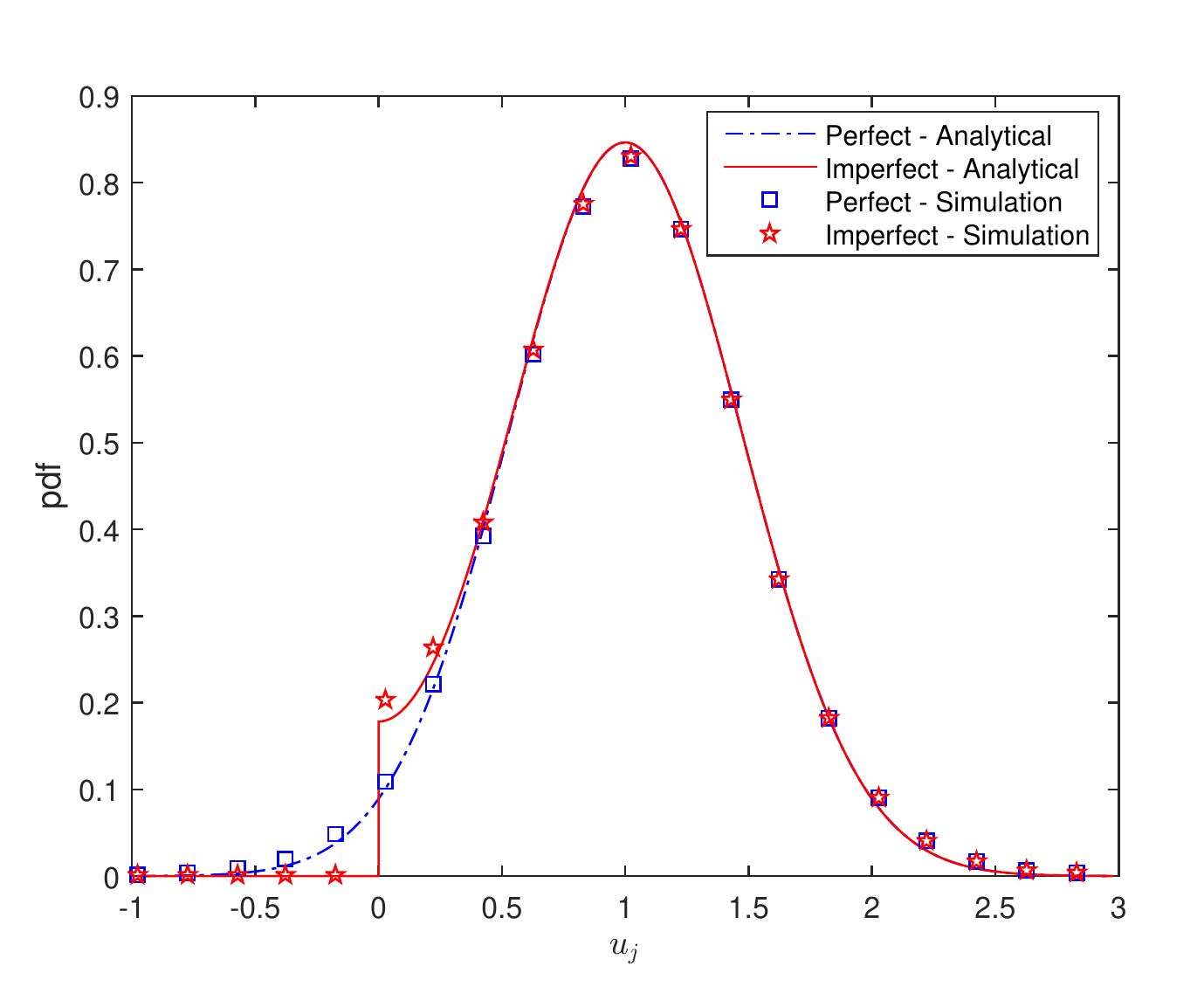}
\label{fig:pdf_u1_n_10_bfsk_0dB}}
%\hfil
\subfloat[$\gamma = 5$~dB]{\includegraphics[width=0.5\textwidth]{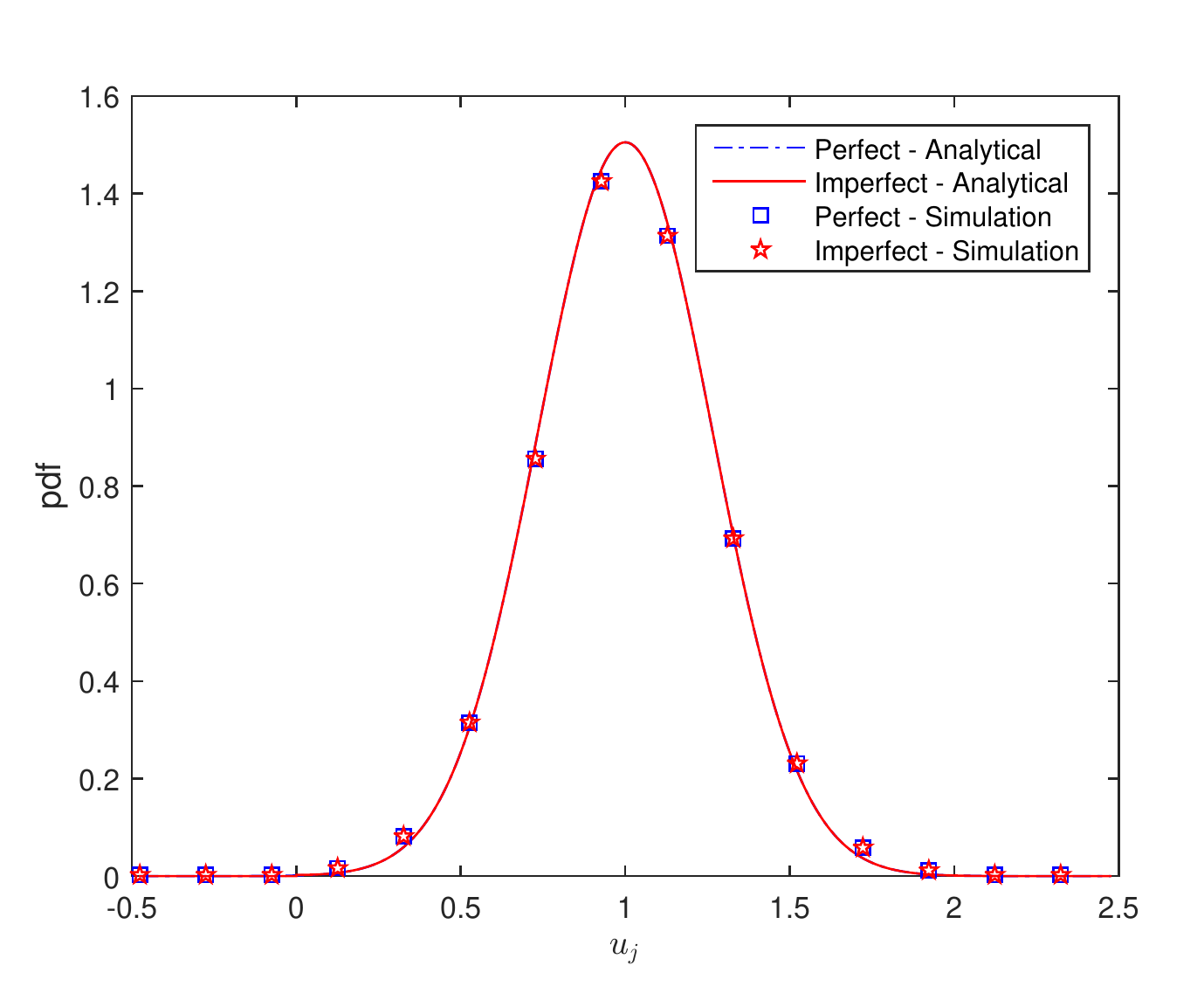}
\label{fig:pdf_u1_n_10_bfsk_5dB}}
\caption{Analytical and simulation results for the pdf of $u_j$ under perfect and imperfect symbol detection cases assuming BFSK modulation and $N_10$.}
\label{fig:pdf_u1_n_10_bfsk}
\end{figure}

The pdf of $u_j$'s under perfect and imperfect detection cases is depicted in Fig.~\ref{fig:pdf_u1_n_10_bpsk}-\ref{fig:pdf_u1_n_10_bfsk} for BPSK and BFSK modulations with the representative alphabets $\mathcal{A}_P=\{1,-1\}$ and $\mathcal{A}_F=\{1,j\}$, respectively. For each modulation, the results are presented at two average SNR values, and the number of receiver antennas are chosen to be $N_10$. We observe that the pdf expressions in Theorem~\ref{theorem:uj_perfect} and Theorem~\ref{theorem:uj_imperfect} match the simulation data successfully for both modulations. The results also verify the approximation for $\bar{\sigma}^2$ given in (\ref{eqn:noise_var_ave_2}), which is the variance of the modified noise $\tilde{v}_{ij}$, under different SNR values and modulations. 

%As is discussed in the analysis of the previous section, the distributions under the imperfect symbol detection case is observed in both Fig.~\ref{fig:pdf_u1_n_10_bpsk} and Fig.~\ref{fig:pdf_u1_n_10_bfsk} to approach or even become identical to the ones associated with the perfect symbol detection for increasing $u_j$ or at moderate to high SNR regime. In Section~\ref{sec:numerical_results}, we show also that any deviations observed in these results between perfect and imperfect detection cases do not have a significant effect on results of outage probabilities.

\subsection{Statistical Analysis of the Ratio $u=u_2/u_1$}\label{sec:u_distribution}
Under various circumstances, (\ref{eqn:binary_ordering}) may be modified to set up a binary ordering rule involving not $u_j$'s individually, but rather their ratio $u = u_2/u_1$. Remembering that $u_j$'s are well-approximated by non-zero Gaussian distribution for both BPSK and BFSK cases, the ratio $u$ can be approximated by the ratio distribution introduced in~\cite{Hinkley69RatioDist}.

\begin{lemma}
Given $u_1$ and $u_2$ be two uncorrelated Gaussian random variables with mean $\mu$ and variance $\sigma^2$, the ratio $u = u_2/u_1$ has the following distribution
\begin{align}\label{eqn:hinkley_exact}
f_u(u) = \frac{c}{\sqrt[]{2\pi}} \frac{1+u}{\left( 1+u^2 \right)^{\frac{3}{2}}} \exp {\left\lbrace -\frac{c^2}{2} \frac{(1-u)^2}{1+u^2}\right\rbrace } \Delta\Phi\left( c \frac{1+u}{\sqrt[]{1+u^2}} \right) + e^{-c^2} f_c(u) \,,
\end{align} 
where $c = \mu/\sigma$, $\Delta\Phi(u) = \Phi(u)-\Phi(-u)$, $\Phi(u)$ is the cdf of the standard Gaussian distribution, and $f_c(u)=\frac{1/\pi}{1+u^2}$ is the pdf of the standard Cauchy distribution.
\end{lemma}
\begin{IEEEproof}
The distribution in (\ref{eqn:hinkley_exact}) is an extension of the one in \cite{Hinkley69RatioDist} for uncorrelated entries, and is rearranged to emphasize the relation to the Cauchy distribution.
\end{IEEEproof}

In order to characterize the distribution of $u$ at low and high SNR, we first employ a $1$st order polynomial to approximate the Gaussian cdf as $\Phi(u) \simeq 0.5\,{+}\, \frac{1}{\sqrt[]{2\pi}} \, u e^{ -u^2/2 }$, which can be obtained via integration by parts~\cite{Gradshteyn} and yields $\Delta\Phi(u) \simeq \frac{2}{\sqrt[]{2\pi}} \, u \, e^{ -u^2/2 }$ with (\ref{eqn:hinkley_exact}) becoming as
\begin{align}\label{eqn:hinkley_low_snr}
f_u(u) \approx f_c(u) \, e^{-c^2} \left[ 1+ c^2 \, \frac{(1+u)^2}{\left( 1+u^2 \right)} \right] .
\end{align}
This result shows that the distribution of the ratio $u$ approaches to the standard Cauchy, i.e., $f_u(u) \rightarrow f_c(u)$, at low SNR for which $c \rightarrow 0$.

At sufficiently large SNR regime for which $c \gg 1$, we have $e^{-c^2} \rightarrow 0$, and the second term in the summation of (\ref{eqn:hinkley_exact}) involving $f_c(u)$ vanishes accordingly. The remaining expression becomes 
\begin{align}\label{eqn:hinkley_high_snr}
f_u(u) \approx \frac{c}{2\,\sqrt[]{\pi}} \, (2-u) \exp \left\lbrace -\frac{c^2}{2}\frac{(1-u)^2}{\left( 1+u^2 \right)} \right\rbrace \,,
\end{align}
where the derivation steps are given in Appendix~\ref{app:fu_high_snr}. Note that, as many communication systems work efficiently under sufficiently large SNR, (\ref{eqn:hinkley_high_snr}) becomes very useful most of the cases.

\begin{figure}[!h]
\centering
\subfloat[BPSK, $\gamma = {-}5$~dB]{\includegraphics[width=0.5\textwidth]{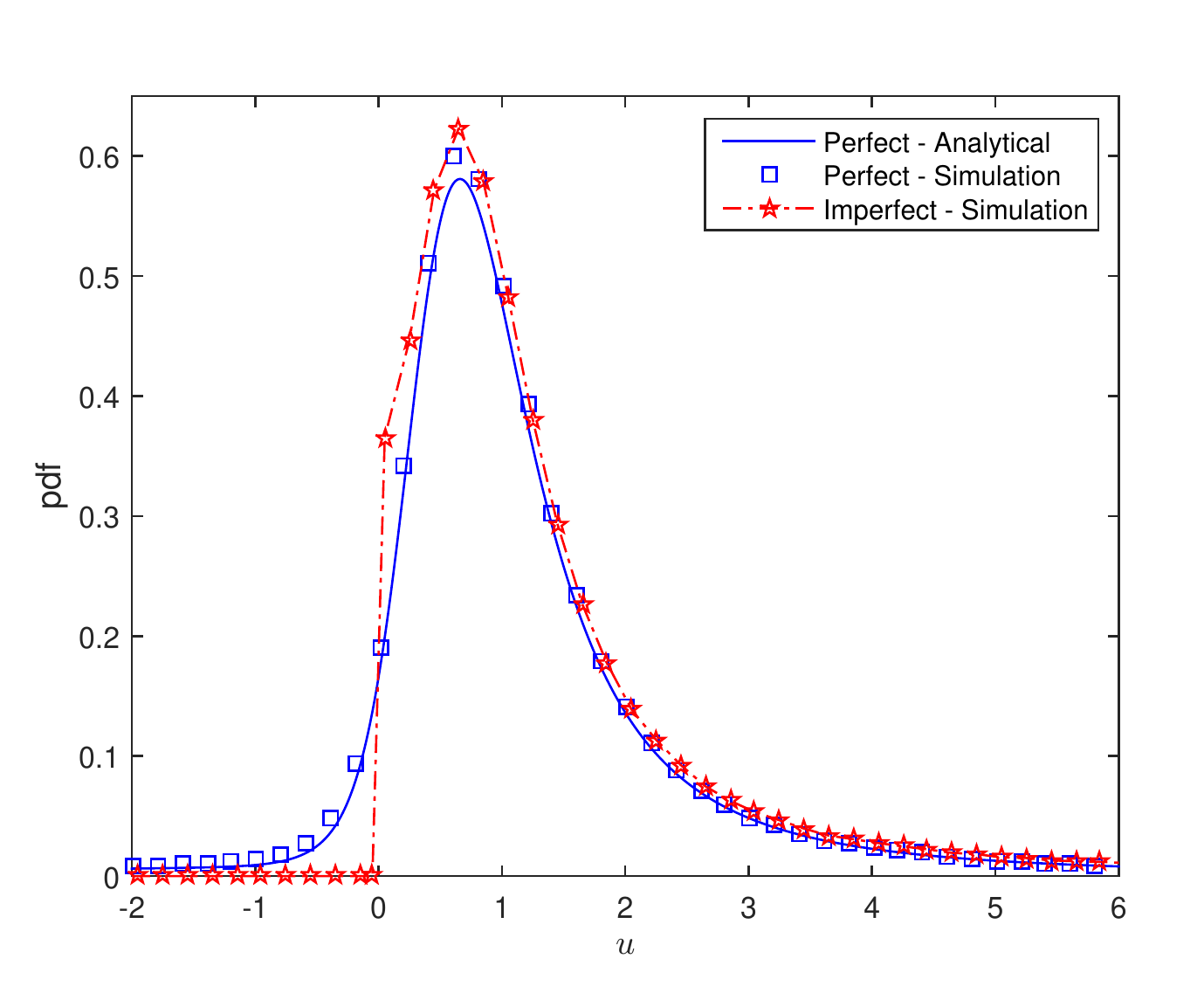}
\label{fig:pdf_u_n_10_bpsk_-5dB}}
\subfloat[BFSK, $\gamma = 5$~dB]{\includegraphics[width=0.5\textwidth]{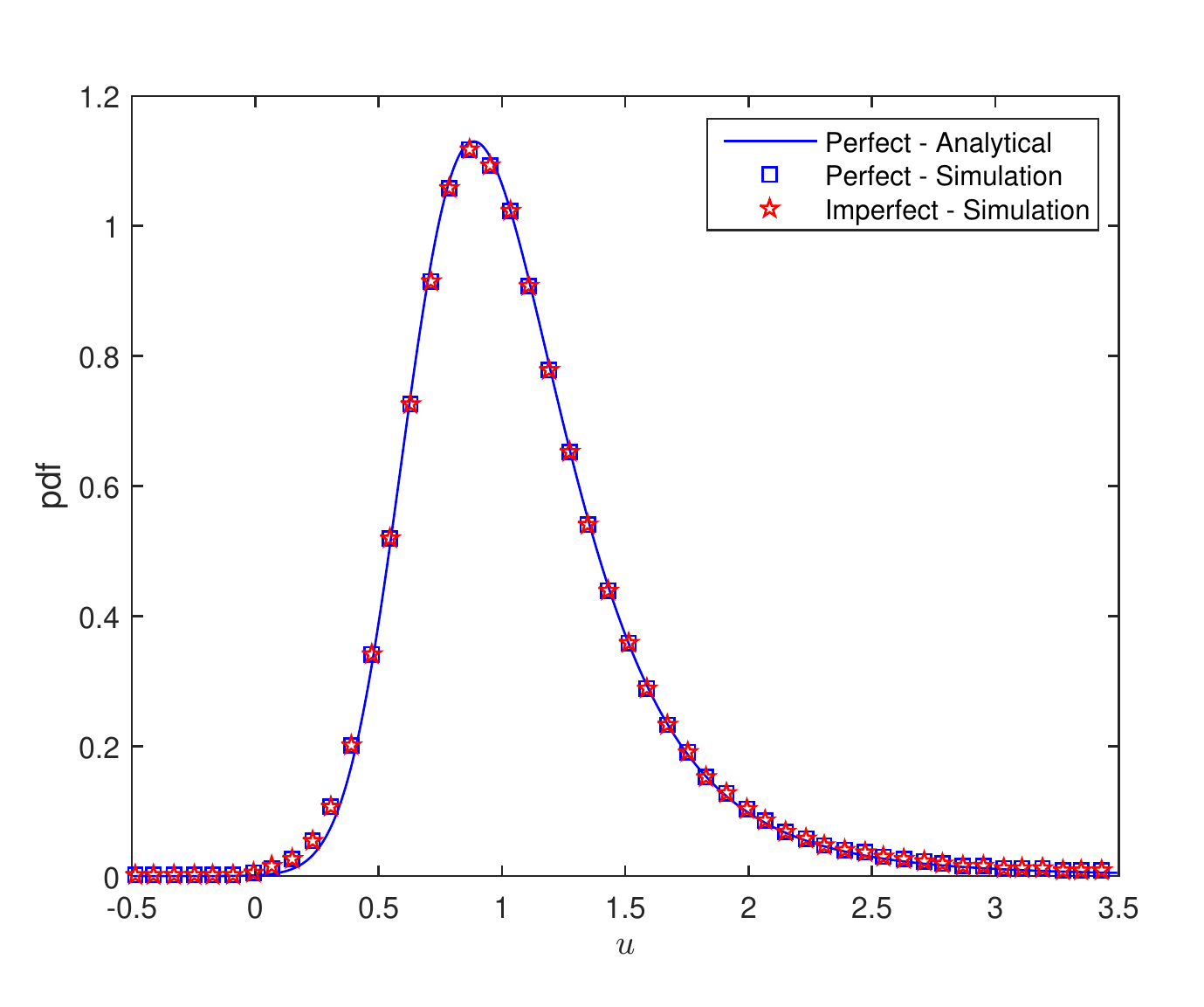}
\label{fig:pdf_u_n_10_bfsk_5dB}}
\caption{Analytical and simulation results for the pdf of $u$ under perfect and imperfect symbol detection cases when $N_10$.}
\label{fig:pdf_u_n_10}
\end{figure}

In Fig.~\ref{fig:pdf_u_n_10}, we depict the numerical results for the pdf of $u$ with the same settings in Section~\ref{sec:statistics_uj}. Without any loss of generality, the results for BPSK and BFSK are presented at $\gamma = -5$~dB and $\gamma = 5$~dB, respectively, which corresponds to the distributions of $u_j$ in Fig.~\ref{fig:pdf_u1_n_10_bpsk}\subref{fig:pdf_u1_n_10_bpsk_-5dB} and Fig.~\ref{fig:pdf_u1_n_10_bfsk}\subref{fig:pdf_u1_n_10_bfsk_5dB}, respectively. We observe that the analytical results successfully match the simulation data for the perfect symbol detection, where almost perfect match is observed in Fig.~\ref{fig:pdf_u_n_10}\subref{fig:pdf_u_n_10_bfsk_5dB} at the moderate SNR of $\gamma = 5$~dB. The deviation between the analytical and simulation data for the imperfect detection is observed to vanish at sufficiently large SNR value of $\gamma = 5$~dB, as expected.

Interestingly, the resulting pdf's depicted in Fig.~\ref{fig:pdf_u_n_10} are not symmetrical as opposed to the constituent random variables $u_j$'s which are Gaussian with perfect symmetry. Indeed, the distribution of $u$ can be approximated with a well-known Cauchy model of symmetrical form as shown in (\ref{eqn:hinkley_low_snr}) which may simplify some analytical results, but it would have poor significance in terms of practical communication systems of interest as this approximation is only valid at low SNR. Note also that pdf of $u$ exhibits relatively long tail characteristics at the low SNR and becomes more compact with a narrow support set at relatively larger SNR.

\section{Outage Analysis over Binary Alphabets}\label{sec:outage_analysis}
In this section, using the statistical behaviour of $u$ derived in Section~\ref{sec:statistical_analysis}, the outage behaviour of M-BLAST is analyzed over binary alphabets with two transmitter antennas. Assuming an $N$-dimensional space $\mathcal{S}$ spanned by the random column vectors of the channel matrix $\mathbf{H}$, i.e., $S=\spn\{\textbf{h}_1,\textbf{h}_2\}$, let $\varphi$ be a random angle between the column vectors in $\mathcal{S}$. Since the norms $\|\textbf{h}_j\|$ have Rayleigh distribution, $\varphi$ is characterized in~\cite{Loyka2004PerAnVB} with the following pdf
\begin{align}\label{eqn:pdf_angle}
f_\varphi (\varphi) = 2\left(N-1\right)\left(\sin\varphi\right)^{2N-3}\cos\varphi \,,
\end{align}
which relies on the discussion on the distribution of the ratio of two random variables with Rayleigh distributions~\cite{Bithas2007Products}~\cite{PapoulisProb}. As a remark for a future study, we may consider correlated MIMO channels with Rician fading to generalize this derivation to mmWave MIMO channels. When Rician fading is assumed, the pdf in \eqref{eqn:pdf_angle} is no more valid and should be derived since the square column norm $\|\textbf{h}_j\|^2$ turns out to have \textit{non-central} chi-square distribution. In addition, the correlated channel assumption requires to derive the joint pdf of correlated square norms $\|\textbf{h}_j\|^2$.

The outage probability at the $1$st detection layer can be evaluated by considering the probability of instantaneous SNR in (\ref{eqn:subchannel_snr}) as follows
\begin{align}\label{eqn:outage1_1}
    F_1(x) &= \Pr \left\lbrace \left\| \textbf{h}_{k_1}^{\bot} \right\|^2 < x \right\rbrace \,,
\end{align}
where the term $E_\textrm{s}/\sigma_{v}^2$ in (\ref{eqn:subchannel_snr}) is dropped as it has no effect on the outage analysis. According to the definition in (\ref{eqn:orth_column}), $\textbf{h}_{j}^{\bot}$ can now be interpreted as the component of $\textbf{h}_{j}$ orthogonal to the other column vector of $\textbf{H}$. Since $\varphi$ is the angle between the column vectors, we have the geometric relation $\|\textbf{h}_j^{\bot}\|^2 = \|\textbf{h}_j\|^2\sin^2\varphi$, and (\ref{eqn:outage1_1}) accordingly becomes
\begin{align}\label{eqn:outage1_2}
F_1(x) &= \int_{0}^{\frac{\pi}{2}} \tilde{F}_1\left(\frac{x}{\sin^2\varphi}\right)  f_\varphi (\varphi) \dd \varphi \,,
\end{align} 
where $\tilde{F}_1(x)$ in the integral expression of (\ref{eqn:outage1_2}) is the probability function given as
\begin{align}
\tilde{F}_1(x) &= \Pr \left\lbrace \left\| \textbf{h}_{k_1} \right\|^2 < x \right\rbrace = \sum\limits_{m{=}1}^{2}\Pr\left\lbrace E_m \right\rbrace \, \Pr \left\lbrace \left\| \textbf{h}_{k_1} \right\|^2 < x \big| E_m \right\rbrace \,, \label{eqn:outage1_4} 
\end{align} 
which follows from the law of total probability where $E_m$ is the ordering event defined in (\ref{eqn:ordering_events}). Based on the fact that $E_m$ is a function of the random variable $u = u_2/u_1$ via (\ref{eqn:binary_ordering}), the outage expression in (\ref{eqn:outage1_4}) is shown in Appendix~\ref{app:outage} to have the following form
\begin{align}\label{eqn:outage1_5}
\tilde{F}_1(x) &= \int_{0}^{\infty} \left[ (1{-}\beta) F_\chi(x) {+} 2\beta \left( F_\chi(x)F_\chi(ux) {-} P\left(u,\sfrac[2pt]{x}{u}\right) \right) \right] f_u(u) \dd u \,,
\end{align}
where $f_u(u)$ is the pdf of the random variable $u$ given in (\ref{eqn:hinkley_exact}), and $\beta = \Pr\{u_1{>}0,u_2{>}0\} - \Pr\{u_1{<}0,u_2{<}0\}$ is a function of joint probabilities. Note that, because $u_j$'s are well-approximated as uncorrelated and Gaussian, we may approximate $\beta$ as follows
\begin{align}
\beta &= \left( \Pr\left\lbrace u_1{>}0 \right\rbrace \right)^2 - \left( \Pr\left\lbrace u_1{<}0 \right\rbrace \right)^2 = 1 - 2 \Phi\left(\frac{u-\mu}{\sigma}\right) \,, \label{eqn:beta}
\end{align}
where $\Phi(x)$ is the cdf of the standard Gaussian distribution, and $\mu$ and $\sigma$ are the mean and standard deviation of $u_j$ derived in Section~\ref{sec:mblast_binary_ordering_rule}.

In (\ref{eqn:outage1_5}), $P(u,a)$ stands for the probability function given as
\begin{align}
P(u,a) & =\int_{0}^{a} F_\chi(uh) \, f_\chi(h) \dd h \label{eqn:pu_1} \\
& = F_\chi(a) - \sum\limits_{r=0}^{N-1} \binom{N{+}r{-}1}{r} \frac{u^r}{(1{+}u)^{N{+}r}} \, \left[ 1 {-} e^{{-}\frac{1{+}u}{2}a} \sum\limits_{i=0}^{N{+}r{-}1} \frac{a^i}{i!} \left(\frac{1{+}u}{2}\right)^{i} \right] \,, \label{eqn:pu_2}
\end{align}
where $F_\chi(x)$ and $f_\chi(x)$ are the cdf and pdf of the chi-square distribution given explicitly in (\ref{eqn:chisquare_cdf}) and (\ref{eqn:chisquare_pdf}), respectively, and $\binom{N{+}r{-}1}{r} = (N{+}r{-}1)!/(N{-}1)!r!$ is the binomial coefficient. Note that, as the upper integral limit in (\ref{eqn:pu_1}) gets larger, i.e., $a\,{\rightarrow}\,\infty$, we have 
\begin{align}
P(u,a\,{\rightarrow}\,\infty) &= 1 - \sum\limits_{r=0}^{n-1} \binom{n{+}r{-}1}{r} \frac{u^r}{(1{+}u)^{n{+}r}} \,, \label{eqn:pu_3}
\end{align}
which can be interpreted also from (\ref{eqn:pu_1}) as the conditional probability $\Pr\{\|\textbf{h}_1\|^2 {<} u\|\textbf{h}_2\|^2 \big| u \}$~\cite{RossIntProbMod}. At moderate to high SNR regime, $P(u,a)$ can be successfully approximated by a $1$st order polynomial, details of which are presented in Appendix~\ref{app:outage}, and the burden in computation of (\ref{eqn:outage1_5}) gets accordingly decreased together with the moderate to high SNR approximation of the pdf of $u$ given in (\ref{eqn:hinkley_high_snr}).

As a final remark, since $u_j\,{=}\,1$ for V-BLAST, we have $u\,{=}\,1$ and hence $\beta\,{=}\,1$, and (\ref{eqn:outage1_5}) accordingly becomes  
\begin{align*}
\tilde{F}_1(x) \big|_{u=1} &= 2 F_\chi^2(x) {-} 2 \int_{0}^{x} F_\chi(h) \, \frac{\partial F_\chi(h)}{\partial h} \dd h \\
&= F_\chi^2(x) \,,
\end{align*} 
which agrees with the result for V-BLAST in~\cite[Eqn.~(21)]{Loyka2004PerAnVB}. 

At the $2$nd detection layer, the M-BLAST algorithm does not need to perform nulling as there are no interfering subchannels. Therefore, the outage probability analysis at the $2$nd layer should consider the column vector itself, $\textbf{h}_{k_2}$, which corresponds to the symbol left undetected at the $1$st layer, as follows 
\begin{align}
F_2(x) &= \Pr \left\lbrace \left\| \textbf{h}_{k_2} \right\|^2 < x \right\rbrace = \sum\limits_{m{=}1}^{2}\Pr\left\lbrace E_m \right\rbrace \, \Pr \left\lbrace \left\| \textbf{h}_{k_2} \right\|^2 < x \, \big| E_m \right\rbrace . \label{eqn:outage2_2}
\end{align}
Following the steps of the $1$st layer, it is shown in Appendix~\ref{app:outage} that (\ref{eqn:outage2_2}) can be expressed as
\begin{align}\label{eqn:outage2_3}
F_2(x) &= \int_{0}^{\infty} \left[ 2\beta \left( -F_\chi(x)F_\chi(ux) {+} P\left(u,\sfrac[2pt]{x}{u}\right) \right) {+} (1{+}\beta) F_\chi(x) \right] f_u(u) \dd u .
\end{align}
In addition, for the case of $\beta\,{=}\,1$ which represents V-BLAST, (\ref{eqn:outage2_3}) becomes 
\begin{align*}
F_2(x) \big|_{u=1} &= -2 F_\chi^2(x) {+} 2 \int_{0}^{x} F_\chi(h) \, \frac{\partial F_\chi(h)}{\partial h} \dd h {+} 2 F_\chi(x) \\
&= - F_\chi^2(x) {+} 2 F_\chi(x) \,,
\end{align*}
which is again the same result obtained in~\cite[Eqn.~(25)]{Loyka2004PerAnVB} for V-BLAST.

\section{Numerical Results}\label{sec:numerical_results}
In this section, we present numerical results regarding error performance and outage probability of M-BLAST in comparison with V-BLAST, in sequence.

\begin{figure}[!p]
\centering
\includegraphics[width=0.66\textwidth]{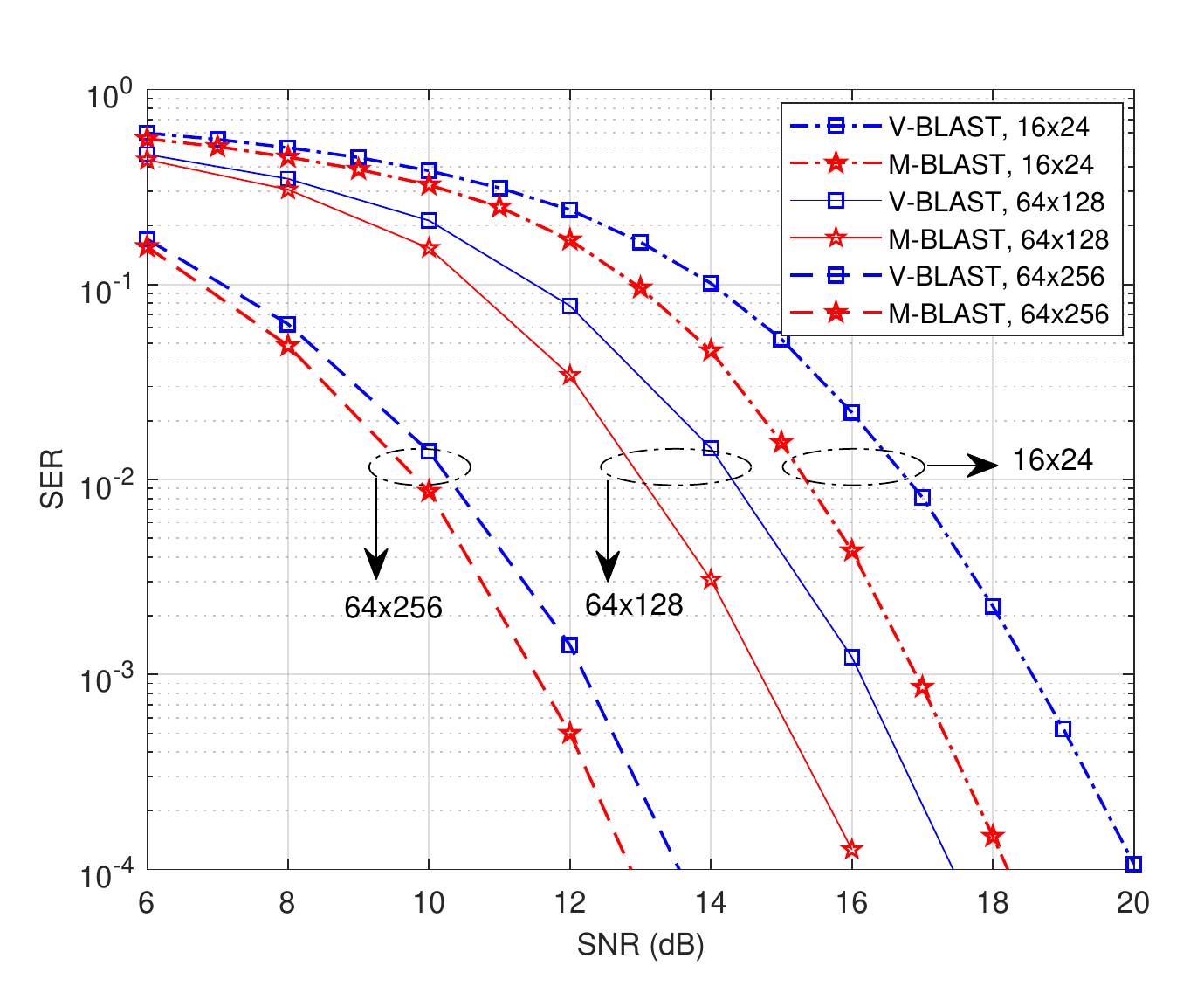}
\caption{SER for $M{\times}N$ MIMO with $(M,N)\,{=}\,\{(16,24),(64,128),(64,256)\}$ antenna pairs with uncorrelated Rayleigh fading channel.}
\label{fig:ser_RayleighMassive_16qam}
\end{figure}

\begin{figure}[!p]
\centering
\includegraphics[width=0.66\textwidth]{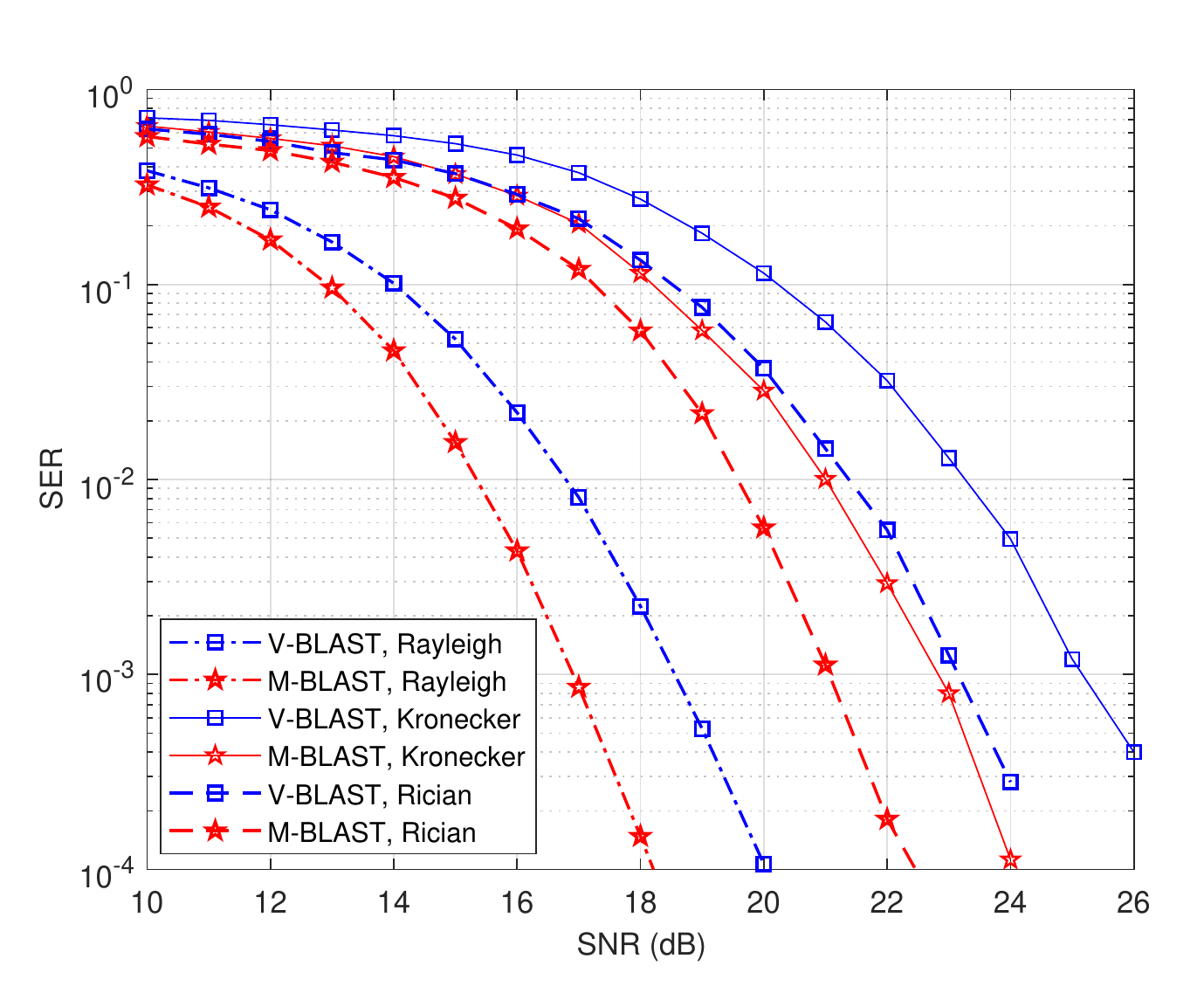}
\caption{SER for $16{\times}24$ MIMO with uncorrelated Rician fading ($K\,{=}\,2$) and Kronecker correlation (exponential transmit and receive correlation matrices with the correlation coefficient of $\rho\,{=}\,0.2$).}
\label{fig:ser_RicianKronecker_16qam}
\end{figure}

\subsection{M-BLAST Error Performance}\label{sec:err_perf}
We first consider the error performance of the M-BLAST and the V-BLAST algorithms over an $M{\times}N$ MIMO system with $(M,N)\,{=}\,\{(16,24),(64,128),(64,256)\}$, where the larger two sets can be considered as a representative setting for the massive MIMO. The modulation is set to be $16$-QAM with the alphabet $\mathcal{A}\,{=}\,\{{\pm}a_1{\pm}j a_2\}$ where $a_1,a_2 \in \{1,3\}$. The Monte Carlo based symbol error rate (SER) results over an uncorrelated Rayleigh fading channel is depicted in Fig.~\ref{fig:ser_RayleighMassive_16qam}, and that for the uncorrelated Rician fading and Kronecker correlation model (exponential transmit and receive correlation matrices with the correlation coefficient of $\rho\,{=}\,0.2$~\cite{Taricco2007OptRec}) is demonstrated in Fig.~\ref{fig:ser_RicianKronecker_16qam}. We observe that M-BLAST has a superior performance over V-BLAST for the various antenna array size, channel fading, and correlation model choices, with as large as $2$~dB SNR improvement. This performance superiority of M-BLAST is very promising since the massive MIMO, correlated MIMO, and Rician fading that we consider in our evaluations reflect the basic characteristics of the next-generation mmWave wireless networks.

\subsection{M-BLAST Outage Performance}
\begin{figure}[!p]
\centering
\subfloat[$F_1(x)$]{\includegraphics[width=0.7\textwidth]{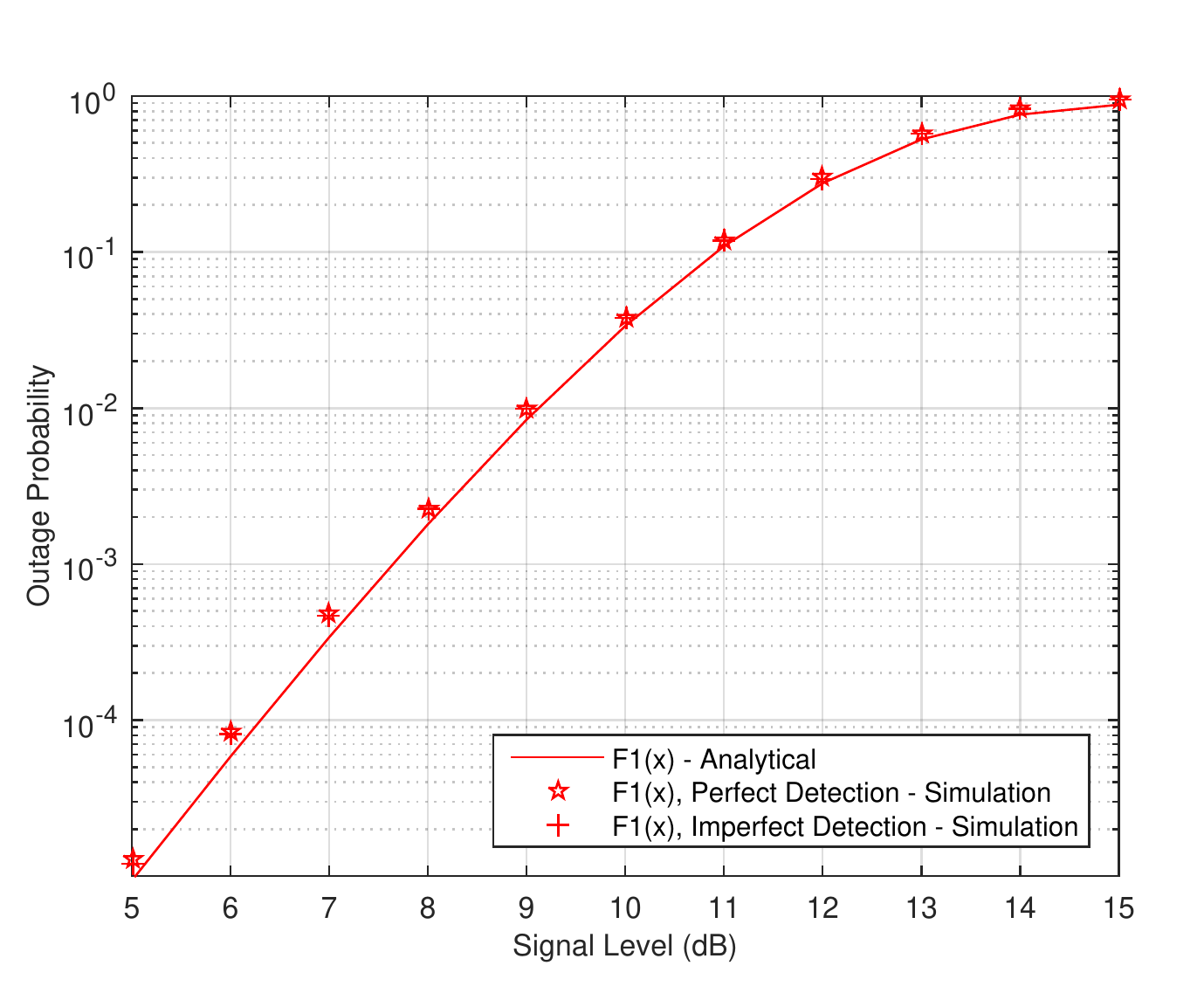}
\label{fig:blast_F1_BPSK_-5dB}}\\
\subfloat[$F_2(x)$]{\includegraphics[width=0.7\textwidth]{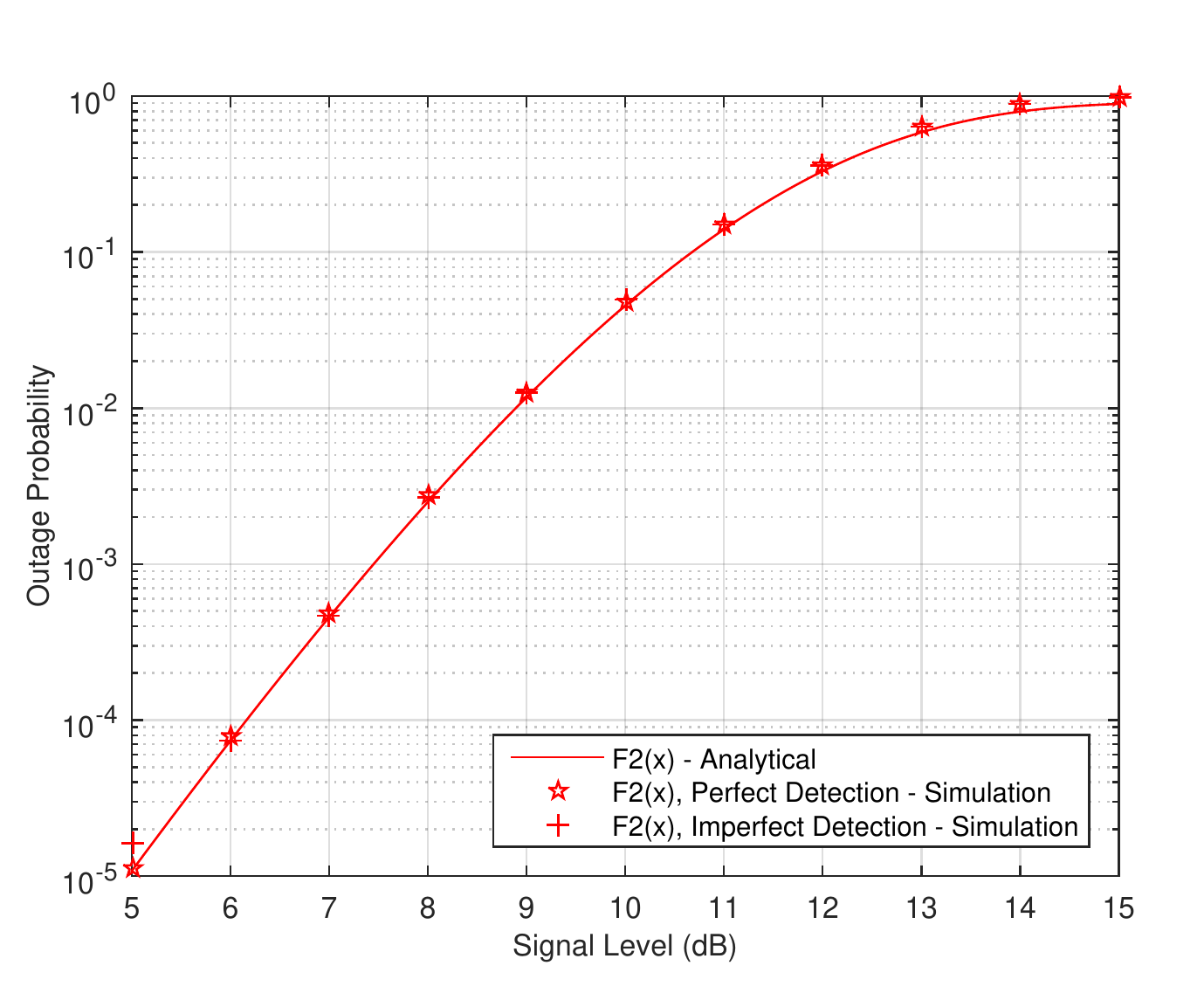}
\label{fig:blast_F2_BPSK_-5dB}}
\caption{Analytical and simulation results for outage probabilities of M-BLAST under BPSK modulation with $N_10$ at $\gamma = {-}5$~dB. Simulation data for perfect and imperfect symbol detection are also provided.}
\label{fig:blast_BPSK_-5dB}
\end{figure}

\begin{figure}[!h]
\centering
\includegraphics[width=0.7\textwidth]{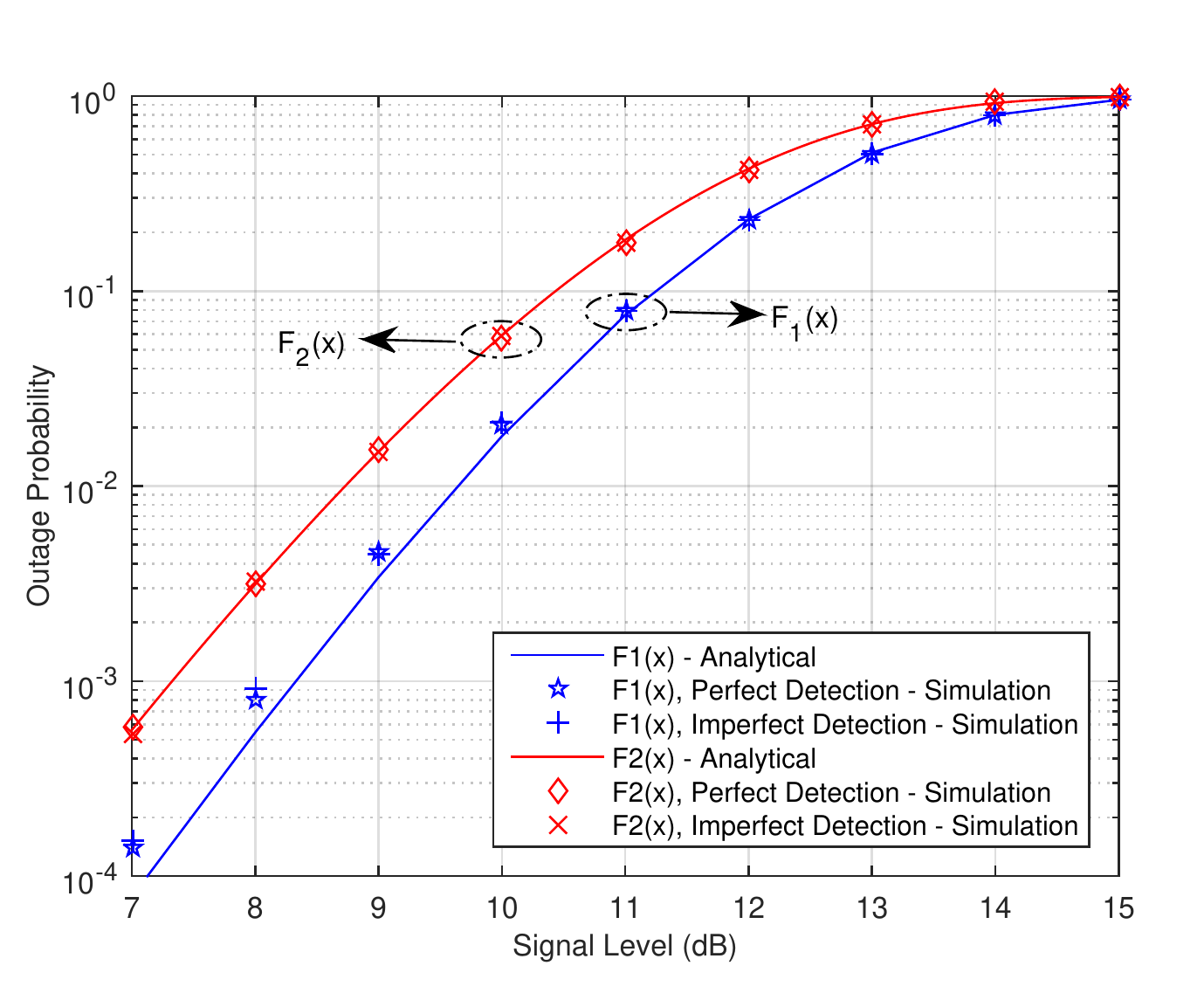}
\caption{Analytical and simulation results for outage probabilities of M-BLAST, $F_1(x)$ and $F_2(x)$, under BFSK modulation with $N_10$ at $\gamma = 5$~dB. Simulation data for perfect and imperfect symbol detection are also provided.}
\label{fig:blast_BFSK_5dB}
\end{figure}

The analytical outage probabilities of M-BLAST at the $1$st and $2$nd detection layers, which are evaluated for $N_10$ according to (\ref{eqn:outage1_2}) and (\ref{eqn:outage2_3}), respectively, are depicted for BPSK in Fig.~\ref{fig:blast_BPSK_-5dB}, and for BFSK in Fig.~\ref{fig:blast_BFSK_5dB}. The results for BPSK and BFSK are evaluated at $\gamma = {-}5$~dB and $\gamma = 5$~dB, respectively, which correspond to the distribution of $u$ given in Fig.~\ref{fig:pdf_u_n_10}\subref{fig:pdf_u_n_10_bpsk_-5dB} and Fig.~\ref{fig:pdf_u_n_10}\subref{fig:pdf_u_n_10_bfsk_5dB}, respectively. The simulation data for M-BLAST is provided under both the perfect and imperfect symbol detection discussed in Section~\ref{sec:mblast_binary_ordering_rule}.

We observe that the analytical results of M-BLAST at both detection layers appear to match the simulation data successfully most of the time under both BPSK and BFSK modulations with different SNR values. Because of the relatively low SNR value considered in Fig.~\ref{fig:blast_BFSK_5dB}, there is no significant difference in outage probabilities of the $1$st and the $2$nd layers, and the associated results are depicted separately to avoid from any confusion while interpreting the results. In addition, although there is a small deviation in the distribution of $u$ between perfect and imperfect symbol detection cases (which appears at a low SNR value of $\gamma={-}5$~dB under BPSK modulation and for small $u$ values shown in Fig.~\ref{fig:pdf_u_n_10}\subref{fig:pdf_u_n_10_bpsk_-5dB}), the resulting outage probabilities are observed not to exhibit a significant difference.

\section{Conclusion}\label{sec:conclusion}
A MAP based symbol detection algorithm, called M-BLAST, is considered over quasi-static MIMO channels with a promise of as large as $2$~dB SNR improvement in various settings of recent interest. The complexity increase as compared to the conventional V-BLAST is linear in the size of modulation alphabet. A rigorous framework is proposed to analyze the performance of the proposed algorithm with binary transmit antennas and binary modulation alphabets. The analytical outage expression for each of the detection layer is derived with a very good match to the simulation data. We believe that the presented analytical framework is very valuable as a contribution to a very limited literature of the performance analysis of ordered SIC receivers, as well as with its potential to be extended to the multiple transmit antennas and multidimensional constellations, as a future work.

\appendices
\section{Proof of Theorem~\ref{theorem:binary_ordering_rule}}\label{app:binary_ordering_rule}
Assuming a binary modulation alphabet $\mathcal{A} = \{a_1,a_2\}$  with $a_1,a_2 \in \mathbb{C}$ and incorporating the density function in (\ref{eqn:likelihood_fnc}), the reliability measure $p_{ij}$ defined in M-BLAST becomes
\begin{align*}
p_{ij} &= \frac{\exp \left\lbrace -\frac{1}{\sigma_{ij}^2} \left\|y_{ij}-s_{ij} \right\|^2 \right\rbrace}{\exp \left\lbrace -\frac{1}{\sigma_{ij}^2} \left\|y_{ij}-a_1 \right\|^2 \right\rbrace + \exp \left\lbrace -\frac{1}{\sigma_{ij}^2} \left\|y_{ij}-a_2 \right\|^2 \right\rbrace} \, ,\\
&= \frac{\exp \left\lbrace \frac{2}{\sigma_{ij}^2} \Real \left( y_{ij} s_{ij}^* \right) \right\rbrace}{\exp \left\lbrace \frac{2}{\sigma_{ij}^2} \Real \left( y_{ij} a_1^* \right) \right\rbrace + \exp \left\lbrace \frac{2}{\sigma_{ij}^2} \Real \left( y_{ij} a_2^* \right) \right\rbrace} \, .
\end{align*}
Defining $\Delta a = a_1 - a_2$ with some manipulations, we have  
\begin{align}
p_{ij} &= \left[ \sum\limits_{m=1}^{2} \exp \left\lbrace \frac{2}{\sigma_{ij}^2} \Real \left( y_{ij} \left(a_m-s_{ij}\right)^* \right) \right\rbrace\right]^{-1} = \left[ 1+ \exp \left\lbrace -\frac{2 b_j}{\sigma_{ij}^2} \Real \left( y_{ij} \Delta a^* \right) \right\rbrace \right]^{-1} \!, \label{eqn:pij_binary_proof}
\end{align} 
where the indicator variable $b_j$ is given as
\begin{equation*}
b_j = \left\{\begin{array}{ll}
+1, & \text{if } s_{ij} = a_1\\
-1, & \text{if } s_{ij} = a_2 
\end{array}\right. \, .
\end{equation*}
Note that the indicator variable can be expressed by using Kronecker delta functions as $b_j=\delta(s_{ij},a_1){-} \delta(s_{ij},a_2)$, which is indeed equal to $\Delta \delta \left( s_{ij} \right)$ in Theorem~\ref{theorem:binary_ordering_rule}. As a result, replacing $\Delta \delta \left( s_{ij} \right)$ with $b_j$ in (\ref{eqn:pij_binary_proof}) yields the probability measures over binary alphabets given in (\ref{eqn:pij_binary}).

When $M\,{=}\,2$, the decision rule for the ordering of subchannels can be fulfilled completely at the $1$st layer by comparing $p_{1j}$'s, which is given after some straightforward steps as follows
\begin{align}\label{eqn:binary_ordering_1}
\frac{1}{\sigma_{11}^2} \Real \left( y_{11} \Delta a^* \right) \Delta \delta \left( s_{11} \right) \gl{E_1}{E_2} 
\frac{1}{\sigma_{12}^2} \Real \left( y_{12} \Delta a^* \right) \Delta \delta \left( s_{12} \right) , 
\end{align}
where $E_m$'s are the ordering events defined in (\ref{eqn:ordering_events}). When (\ref{eqn:noise_var}) is also incorporated for $\sigma_{1j}^2$'s in (\ref{eqn:binary_ordering_1}), we have
\begin{align}\label{eqn:binary_ordering_2}
\left\| \textbf{h}_1^{\bot} \right\|^2 \Real\left\lbrace \, y_{11} \, \Delta a^* \, \right\rbrace \Delta \delta \left( s_{11} \right)
\gl{E_1}{E_2} \; 
\left\| \textbf{h}_2^{\bot} \right\|^2 \Real\left\lbrace \, y_{12} \, \Delta a^* \, \right\rbrace \Delta \delta \left( s_{12} \right) \, .
\end{align}
Employing the relation $\|\textbf{h}_j^{\bot}\|^2\,{=}\,\|\textbf{h}_j\|^2\sin^2\varphi$ in (\ref{eqn:binary_ordering_2}), which is a result of the discussion Section~\ref{sec:outage_analysis}, cancelling the terms $\sin^2\varphi$ from both sides, and defining $u_j\,{=}\,\Real\{y_{1j}\Delta a^*\}\Delta\delta(s_{1j})$ as in (\ref{eqn:uj}), we obtain (\ref{eqn:binary_ordering}).  

\section{Proof of Theorem~\ref{theorem:uj_perfect}}\label{app:distribution_uj_perfect}
Assuming perfect symbol detection such that $s_{1j}\,{=}\,x_j$ for $j\,{=}\,1,2$, the random variable $u_j$ given in (\ref{eqn:uj_bpsk}) for BPSK can be expressed by employing (\ref{eqn:subchannel_defn}) as 
\begin{align*}
u_j &= \Real \left\lbrace y_{1j} \right\rbrace \sgn \left( x_j \right) = a_1 + \sgn \left( x_j \right) \Real \left\lbrace \, \tilde{v}_{1j} \right\rbrace \, ,
\end{align*}
which makes use of the fact $x_{j} \sgn \left( x_j \right)\,{=}\,a_1$ for $\forall x_j \in \mathcal{A}_P$. Because $\sgn(x_j)^2\,{=}\,1$ for $\forall x_j \in \mathcal{A}_P$, $u_{j}$ can be approximated as real-valued Gaussian with mean $a_1$ and variance $\bar{\sigma}^2\!/2$, i.e., $u_{j}\,{\sim}\,\mathcal{N}\left(a_1,\bar{\sigma}^2\!/2 \right)$, where the associated pdf is given as
\begin{align}\label{eqn:pdf_per_bpsk}
f_{\mathcal{N}}\left(x,a_1,\bar{\sigma}^2\!/2\right) &= \frac{1}{\sqrt[]{\pi\bar{\sigma}^2}} \exp\left\lbrace{-}\frac{(x{-}a_1)^2}{\bar{\sigma}^2} \right\rbrace \, .
\end{align}
In addition, the covariance of $u_j$'s is given as  
\begin{align*}
\cov (u_1,u_2) &= \mathbb{E}\{ (u_1-a_1)(u_2-a_1)\} = \left( \mathbb{E} \left\lbrace \sgn \left( x_j \right) \right\rbrace \right)^2 \, \mathbb{E}\left\lbrace \Real \left\lbrace \, \tilde{v}_{11} \right\rbrace \Real \left\lbrace \, \tilde{v}_{12} \right\rbrace \right\rbrace = 0 \, ,
\end{align*}
which reveals that $u_j$'s are also uncorrelated.

Similarly, $u_j$ in (\ref{eqn:uj_bfsk}) for BFSK is given as
\begin{align}
u_j &= \Real\left\lbrace y_{1j} (1+j) \right\rbrace \frac{x_j^2}{a_1} = \Delta y_{1j} \frac{x_j^2}{a_1} \, , \label{eqn:per_bfsk_1}
\end{align}
where $\Delta y_{1j}\,{=}\,\Real \left\lbrace y_{1j} \right\rbrace {-} \Imag \left\lbrace y_{1j} \right\rbrace$ is defined to be
\begin{equation}\label{eqn:per_bfsk_2}
\Delta y_{1j} = \Bigg\{\!\!
  \begin{array}{rl}
    a_1    {+} \Delta \tilde{v}_{1j} \, , \;\; & \text{if} \;\; x_j\,{=}\,a_1 \\
    {-}a_1 {+} \Delta \tilde{v}_{1j} \, , \;\; & \text{if} \;\; x_j\,{=}\,j a_1 
  \end{array} \, ,
\end{equation}
and is formulated as $\Delta y_{1j}\,{=}\,\Delta \tilde{v}_{1j} + x_j^2/a_1$ for $\forall x_j \in \mathcal{A}_F$. In (\ref{eqn:per_bfsk_2}), $\Delta \tilde{v}_{1j}\,{=}\,\Real \left\lbrace \tilde{v}_{1j} \right\rbrace {-} \Imag \left\lbrace \tilde{v}_{1j} \right\rbrace$ is real-valued Gaussian with zero-mean and $\bar{\sigma}^2$ variance, i.e., $\Delta \tilde{v}_{1j} \,{\sim}\, \mathcal{N}\left(0,\bar{\sigma}^2 \right)$. Employing the compact expression of $\Delta y_{1j}$ in (\ref{eqn:per_bfsk_1}) gives
\begin{align}
u_j &= a_1^2 + \frac{x_j^2}{a_1} \Delta \tilde{v}_{1j} \, ,
\end{align}
which employs the fact $x_j^4\,{=}\,a_1^4$ for $\forall x_j \in \mathcal{A}_F$, and shows that $u_{j}$ has a real-valued Gaussian distribution with mean $a_1^2$ and variance $a_1^2\bar{\sigma}^2$, i.e., $u_{j}\,{\sim}\,\mathcal{N}\left(a_1^2,a_1^2\bar{\sigma}^2 \right)$, where the associated pdf is
\begin{align}\label{eqn:pdf_per_bfsk}
f_{\mathcal{N}}\left(x,a_1^2,a_1^2\bar{\sigma}^2\right) &= \frac{1}{\sqrt[]{2\pi(a_1\bar{\sigma})^2}} \exp\left\lbrace{-}\frac{(x{-}a^2_1)^2}{2(a_1\bar{\sigma})^2} \right\rbrace .
\end{align}
In addition, the covariance of $u_j$'s is given as  
\begin{align*}
\cov (u_1,u_2) &= \mathbb{E}\{ (u_1-a_1^2)(u_2-a_1^2)\} = \frac{1}{a_1^2} \left( \mathbb{E} \left\lbrace x_j^2 \right\rbrace \right)^2 \, \mathbb{E} \left\lbrace \Delta \tilde{v}_{11} \Delta \tilde{v}_{12} \right\rbrace = 0 \, , 
\end{align*}
which follows from the fact that $\mathbb{E} \left\lbrace x_j^2 \right\rbrace\,{=}\,0$ over the alphabet $\mathcal{A}_F$, and shows that $u_j$'s are uncorrelated, as is the case for BPSK.

\section{Proof of Theorem~\ref{theorem:uj_imperfect}}\label{app:distribution_uj_imperfect}
When the minimum distance receiver for BPSK modulation with the alphabet $\mathcal{A}_P =\{a_1,{-}a_1\}$ is assumed, the detection rule is given as $s_{1j}=a_1\sgn(\Real(y_{1j}))$, and $u_j$ in (\ref{eqn:uj_bpsk}) becomes
\begin{align}
u_j &= \Real \left\lbrace y_{1j} \right\rbrace \sgn \left( \Real \left( y_{1j} \right) \right) = \left( x_j {+} w_j \right) \, \sgn \left( x_j {+} w_j \right) \,,\label{eqn:imper_bpsk_1}  
\end{align}
where $w_j=\Real(\tilde{v}_{1j})$. In order to identify the distribution of $u_j$ given in (\ref{eqn:imper_bpsk_1}) for BPSK modulation with the minimum distance receiver, consider the following probability
\begin{align}
\Pr\left\lbrace u_j {<} x | x_j \right\rbrace &= \Pr\left\lbrace x_j {+} w_j {<} x , \, x_j {+} w_j {>} 0 \right\rbrace + \Pr\left\lbrace x_j {+} w_j {>} {-}x , \, x_j {+} w_j {<} 0 \right\rbrace \,,\nonumber \\  
&= \Pr\left\lbrace 0 {<} x_j {+} w_j {<} x \right\rbrace + \Pr\left\lbrace {-}x {<} x_j {+} w_j{<}0 \right\rbrace \,,\nonumber \\
&= F_w(x{-}x_j){-}F_w({-}x{-}x_j) \,,\label{eqn:imper_bpsk_2}
\end{align}
for $x\,{\geq}\,0$, and where $F_w(x)$ is the cdf (cumulative distribution function) of $w_j$. Taking average of (\ref{eqn:imper_bpsk_2}) over the alphabet $\mathcal{A}_P$ gives the cdf of $u_j$ as follows
\begin{align}
\Pr\left\lbrace u_j {<} x \right\rbrace &= \frac{1}{2} \left[ F_w(x{-}a_1){-}F_w({-}x{-}a_1){+}F_w(x{+}a_1){-}F_w({-}x{+}a_1) \right] \,,\nonumber\\
&= {-}1{+}F_w(x{+}a_1){+}F_w(x{-}a_1) \,,\nonumber\\
&= {-}1{+}\frac{1}{\sqrt[]{\pi\bar{\sigma}^2}} \int_{-\infty}^{x} \left[ \exp\left\lbrace {-}\frac{(w{-}a_1)^2}{\bar{\sigma}^2}\right\rbrace {+} \exp\left\lbrace {-}\frac{(w{-}a_1)^2}{\bar{\sigma}^2}\right\rbrace\right] \dd w . \label{eqn:imper_bpsk_3} 
\end{align}
The desired pdf is accordingly obtained by taking derivative of (\ref{eqn:imper_bpsk_3}) with respect to $x$, which yields
\begin{align}
f_{u_{j}|\textrm{P}} (x) &= \frac{1}{\sqrt[]{\pi\bar{\sigma}^2}} \left[ \exp\left\lbrace {-}\frac{(x{+}a_1)^2}{\bar{\sigma}^2} \right\rbrace {+} \exp \left\lbrace {-}\frac{(x{-}a_1)^2}{\bar{\sigma}^2} \right\rbrace \right] \,, \label{eqn:imper_bpsk_4}
\end{align}
where the desired expression of (\ref{eqn:pdf_imper_bpsk}) is obtained by replacing (\ref{eqn:pdf_per_bpsk}) in (\ref{eqn:imper_bpsk_4}).

For BFSK modulation with the alphabet $\mathcal{A}_p =\{a_1,j a_1\}$, the minimum distance detector is given as
\begin{equation}\label{eqn:imper_bfsk_1}
s_{1j} = \Bigg\{\!
  \begin{array}{rl}
    a_1,   \;\; & \text{if} \;\; \Delta y_{1j}{>}0 \\
    j a_1, \;\; & \text{if} \;\;  \Delta y_{1j}{<}0
  \end{array} \,,
\end{equation}
where $\Delta y_{1j}\,{=}\,\Real(y_{1j}){-}\Imag(y_{1j})$, and (\ref{eqn:uj_bfsk}) accordingly becomes $u_j\,{=}\,a_1\Delta y_{1j}\sgn(\Delta y_{1j})$. In order to characterize the cdf of $u_j$, consider the following probability 
\begin{align}
\Pr\left\lbrace u_j {<} x \right\rbrace &= \Pr\left\lbrace a_1 \Delta y_{1j} {<} x , \, \Delta y_{1j} {>} 0 \right\rbrace + \Pr\left\lbrace {-}a_1 \Delta y_{1j} {<} x , \, \Delta y_{1j}{<}0 \right\rbrace \,,\nonumber \\  
&= \Pr\left\lbrace {-}\frac{x}{a_1} {<} \Delta y_{1j} {<} \frac{x}{a_1} \right\rbrace  \,,\label{eqn:imper_bfsk_2}
\end{align}
for $x\,{\geq}\,0$. Employing (\ref{eqn:per_bfsk_2}) in (\ref{eqn:imper_bfsk_2}) via the law of total probability gives the cdf of $u_j$ as follows 
\begin{align}
\Pr\left\lbrace u_j {<} x \right\rbrace &= \frac{1}{2} \left[ \Pr\left\lbrace {-}\frac{x}{a_1} {<} a_1{+}\Delta\tilde{v}_{1j} {<} \frac{x}{a_1} \right\rbrace {+}  \Pr\left\lbrace {-}\frac{x}{a_1} {<} {-}a_1{+}\Delta \tilde{v}_{1j} {<} \frac{x}{a_1} \right\rbrace \right] \,,\nonumber\\
&= {-}1{+}F_{\tilde{v}}\left({-}a_1{+}\frac{x}{a_1}\right){+}F_{\tilde{v}}\left(a_1{+}\frac{x}{a_1}\right) \,, \nonumber\\
&= {-}1{+}\frac{1}{\sqrt[]{2\pi (a_1\bar{\sigma})^2}} \int_{-\infty}^{x} \left[ \exp\left\lbrace {-}\frac{(w{-}a_1^2)^2}{2(a_1\bar{\sigma})^2}\right\rbrace {+} \exp\left\lbrace {-}\frac{(w{+}a_1^2)^2}{2(a_1\bar{\sigma})^2}\right\rbrace\right] \dd w \,, \label{eqn:imper_bfsk_4}
\end{align}
where $F_{\tilde{v}}(x)$ is the cdf of $\Delta \tilde{v}_{1j}$. The pdf is then obtained by taking derivative of (\ref{eqn:imper_bfsk_4}), which produces
\begin{align}
f_{u_{j}|\textrm{F}} (x) &= \frac{1}{\sqrt[]{2\pi (a_1\bar{\sigma})^2}} \left[ \exp\left\lbrace {-}\frac{(x{-}a_1^2)^2}{2(a_1\bar{\sigma})^2}\right\rbrace {+} \exp\left\lbrace {-}\frac{(x{+}a_1^2)^2}{2(a_1\bar{\sigma})^2}\right\rbrace\right] \,,
\label{eqn:imper_bfsk_5}  
\end{align}
where the pdf expression in (\ref{eqn:pdf_imper_bfsk}) is obtained by replacing (\ref{eqn:pdf_per_bfsk}) in (\ref{eqn:imper_bfsk_5}).

\section{PDF of $u$ at Moderate to High SNR}\label{app:fu_high_snr}
Because the variance $\sigma^2$, which is indeed a multiple of $\bar{\sigma}^2$ for both BPSK and BFSK, is inversely proportional to the input SNR according to (\ref{eqn:noise_var_ave_2}), the support set for which the distribution function $f_u(u)$ takes non-zero values is confined to a narrow interval $\mathcal{U}_s=[ 1-\epsilon_1,1+\epsilon_2 ]$ as SNR gets larger, where $\epsilon_i \ll 1$ for $i = 1,2$. Observing the nonlinear rational polynomial $p(u)$ in the first term of the summation in (\ref{eqn:hinkley_exact}) changes slowly over $\mathcal{U}_s$, it can be approximated by a linear function around $u\,{=}\,1$ as follows  
\begin{align*}
p(u) = \frac{1+u}{\left( 1+u^2 \right)^{\frac{3}{2}}} &\approx m_p \, (u-1) + p(1) \,,
\end{align*}
where $p(1)\,{=}\,1/\sqrt[]{2}$, and the slope $m_p$ of $p(u)$ at $u\,{=}\,1$ is
\begin{align*}
m_p &= \left. \frac{\partial\, p(u)}{\partial\, u} \right|_{u=1} 
= \left. \frac{1-3u-2u^2}{\left( 1+u^2\right)^{5/2}} \right|_{u=1} = \frac{-1}{\sqrt[]{2}} \,,
\end{align*}
which yields
\begin{align}\label{eqn:hinkley_high_snr_approx1}
p(u) &\approx \frac{-u+2}{\sqrt[]{2}} . 
\end{align}

In addition, because $c\,{\gg}\,1$ at high SNR, and $1{+}u\,{>}\,0$ for $u\,{\in}\,\mathcal{U}_s$, the term involving the function $\Delta\Phi$ in (\ref{eqn:hinkley_exact}) can be approximated as follows
\begin{align}\label{eqn:hinkley_high_snr_approx2}
\Delta\Phi\left( c \frac{1+u}{\sqrt[]{1+u^2}} \right) &= \Phi\left( c \frac{1+u}{\sqrt[]{1+u^2}} \right) - \Phi\left( -c \frac{1+u}{\sqrt[]{1+u^2}} \right) \approx 1 \,,
\end{align}
which directly follows from the fact that the $\Phi(x) \approx 1$ and $\Phi(-x) \approx 0$ for $x \gg 0$, by definition, where $x$ represents the argument of the function $\Delta\Phi$. As a result, employing (\ref{eqn:hinkley_high_snr_approx1}) and (\ref{eqn:hinkley_high_snr_approx2}) in (\ref{eqn:hinkley_exact}) yields the result in (\ref{eqn:hinkley_high_snr}). Note that the exponential term in (\ref{eqn:hinkley_high_snr}) can be further simplified by using Taylor's expansion, but, which comes with a certain amount of deviation over tails, i.e., $u \rightarrow \left\lbrace 1-\epsilon_1,1+\epsilon_2 \right\rbrace$, depending on the polynomial degree used in the expansion.

\section{Derivation of $F_1(x)$ and $F_2(x)$}\label{app:outage}
At the $1$st layer, consider the outage expression $\tilde{F}_1(x)$ in (\ref{eqn:outage1_4}) which can be rearranged as  
\begin{align}
\tilde{F}_1(x) &= \Pr\left\lbrace\left\|\textbf{h}_1\right\|^2 < x , \, E_1 \right\rbrace + \Pr\left\lbrace\left\|\textbf{h}_2 \right\|^2 < x , \, E_2 \right\rbrace \,, \label{eqn:app:outage_1} 
\end{align}
where the first term can be further expanded for a given $u$ by employing the law of total probability as follows
\begin{align}
\Pr\left\lbrace\left\|\textbf{h}_1\right\|^2 {<} \, x , E_1 \big| u \right\rbrace &= \Pr\left\lbrace\left\|\textbf{h}_1\right\|^2 {<}\,x , \left\|\textbf{h}_{1}\right\|^2 u_1 \, {>} \left\|\textbf{h}_{2}\right\|^2 u_2 \right\rbrace \label{eqn:app:outage_2}\\
&= \Pr\left\lbrace\left\|\textbf{h}_1\right\|^2 {<} \, x , \left\|\textbf{h}_{1}\right\|^2 {>}\,u \left\|\textbf{h}_{2}\right\|^2 \big|\,u\,{>}\,0 \right\rbrace \Pr\left\lbrace u_1{>}0,u_2{>}0 \right\rbrace \label{eqn:app:outage_3}\\ 
& \; + \Pr\left\lbrace\left\|\textbf{h}_1\right\|^2 {<} \, x , \left\|\textbf{h}_{1}\right\|^2 {>}\,u \left\|\textbf{h}_{2}\right\|^2 \big|\,u\,{<}\,0 \right\rbrace \Pr\left\lbrace u_1{>}0,u_2{<}0 \right\rbrace \label{eqn:app:outage_4}\\
& \; + \Pr\left\lbrace\left\|\textbf{h}_1\right\|^2 {<} \, x , \left\|\textbf{h}_{1}\right\|^2 {<}\,u \left\|\textbf{h}_{2}\right\|^2 \big|\,u\,{<}\,0 \right\rbrace \Pr\left\lbrace u_1{<}0,u_2{>}0 \right\rbrace \label{eqn:app:outage_5}\\
& \; + \Pr\left\lbrace\left\|\textbf{h}_1\right\|^2 {<} \, x , \left\|\textbf{h}_{1}\right\|^2 {<}\,u \left\|\textbf{h}_{2}\right\|^2 \big|\,u\,{>}\,0 \right\rbrace \Pr\left\lbrace u_1{<}0,u_2{<}0 \right\rbrace . \label{eqn:app:outage_6}
\end{align}
Because $\|\textbf{h}_j\|^2\,{\geq}\,0$ for any realization, the probability on the left in (\ref{eqn:app:outage_4}) becomes $\Pr\{\|\textbf{h}_1\|^2\,{<}\,x\}$, and (\ref{eqn:app:outage_5}) simply vanishes. Furthermore, combining (\ref{eqn:app:outage_3}) and (\ref{eqn:app:outage_6}) produces the expression
\begin{align}
\Pr\left\lbrace\left\|\textbf{h}_1\right\|^2 {<}\,x , E_1 \big|\,u \right\rbrace &= \alpha_1 \Pr\left\lbrace\left\|\textbf{h}_1\right\|^2 {<}\,x \right\rbrace + \beta \Pr\left\lbrace\left\|\textbf{h}_1\right\|^2 {<}\,x , \left\|\textbf{h}_{1}\right\|^2 {>}\,u \left\|\textbf{h}_{2}\right\|^2 \big|\,u\,{>}\,0 \right\rbrace \,,\label{eqn:app:outage_8} 
\end{align}
where $\alpha_1 = \Pr\{u_1{>}0,u_2{<}0\} + \Pr\{u_1{<}0,u_2{<}0\}$ and $\beta = \Pr\{u_1{>}0,u_2{>}0\} - \Pr\{u_1{<}0,u_2{<}0\}$. Integrating over the distribution of $\|\textbf{h}_2\|^2$ for $u\,{\geq}\,0$, (\ref{eqn:app:outage_8}) can be further elaborated  as follows
\begin{align}
\Pr\left\lbrace\left\|\textbf{h}_1\right\|^2 {<}\,x , E_1 \big| u \right\rbrace &= \alpha_1 F_\chi(x) + \beta \int_{0}^{\infty} \Pr\left\lbrace\left\|\textbf{h}_1\right\|^2 {<}\,x , \left\|\textbf{h}_{1}\right\|^2 {>}\,uh \right\rbrace f_\chi(h) \dd h  \,, \nonumber \\
&= \alpha_1 F_\chi(x) + \beta \int_{0}^{\frac{x}{u}} \Pr\left\lbrace uh\,{<} \left\|\textbf{h}_1\right\|^2 {<}\,x \right\rbrace f_\chi(h) \dd h \,, \nonumber \\
&= \alpha_1 F_\chi(x) + \beta \int_{0}^{\frac{x}{u}} \left( F_\chi(x) - F_\chi(uh) \right) f_\chi(h) \dd h\nonumber \,, \\ 
&= \alpha_1 F_\chi(x) + \beta \left[ F_\chi(x)F_\chi\left(\sfrac[2pt]{x}{u}\right) - \int_{0}^{\frac{x}{u}} \! F_\chi(uh) f_\chi(h) \dd h \right] . \label{eqn:app:outage_12}  
\end{align}

The second term in (\ref{eqn:app:outage_1}) can be evaluated in a similar way as follows
\begin{align}
\Pr\left\lbrace\left\|\textbf{h}_2\right\|^2 {<}\,x , E_2 \big| u \right\rbrace &= \Pr\left\lbrace\left\|\textbf{h}_2\right\|^2 {<}\,x ,  \left\|\textbf{h}_{1}\right\|^2 u_1\,{<} \left\|\textbf{h}_{2}\right\|^2 u_2 \right\rbrace \label{eqn:app:outage_13}\\
&= \alpha_2 F_\chi(x) + \beta \left[ F_\chi(x)F_\chi(ux) - \int_{0}^{ux} \!\!\! F_\chi\left(\sfrac[2pt]{h}{u}\right) f_\chi(h) \dd h \right] \,, \label{eqn:app:outage_14}
\end{align}
where $\alpha_2 = \Pr\{u_1{<}0,u_2{>}0\} + \Pr\{u_1{<}0,u_2{<}0\}$. Realizing the similarity between (\ref{eqn:app:outage_2}) and (\ref{eqn:app:outage_13}), the result in (\ref{eqn:app:outage_14}) follows directly from (\ref{eqn:app:outage_12}) by replacing $u$ with $1/u$. Note that, because $u$ and $1/u$ have identical distributions as $u_1$ and $u_2$ are distributed identically, as well, they can be used interchangeably for each term of the summations in (\ref{eqn:app:outage_12}) and (\ref{eqn:app:outage_14}) without making any difference in the integration over $u$ given in (\ref{eqn:outage1_5}). Based on this discussion and realizing that $\alpha_1{+}\alpha_2=1{-}\beta$, adding up the terms in (\ref{eqn:app:outage_12}) and (\ref{eqn:app:outage_14}) gives the outage expression
\begin{align}
\tilde{F}_1(x|u) &= (1-\beta) F_\chi(x) + 2 \beta \left[ F_\chi(x)F_\chi(ux) - P\left(u,\sfrac[2pt]{x}{u}\right) \right] \,, \label{eqn:app:outage_15}
\end{align}
which achieves (\ref{eqn:outage1_4}) after integration over the distribution of $u$ with $u\,{\geq}\,0$, and where the probability function $P(u,a)$ is defined in (\ref{eqn:pu_1}).

At the $2$nd layer, the outage expression in (\ref{eqn:outage2_2}) can be given as
\begin{align}
F_2(x) &= \Pr\left\lbrace\left\|\textbf{h}_2\right\|^2 < x , \, E_1 \right\rbrace + \Pr\left\lbrace\left\|\textbf{h}_1 \right\|^2 < x , \, E_2 \right\rbrace, \label{eqn:app:out2_1} 
\end{align}
where the first term can be expanded for a given $u$ as follows
\begin{align}
\Pr\left\lbrace\left\|\textbf{h}_2\right\|^2 {<}\,x , E_1 \big| u \right\rbrace &= \Pr\left\lbrace\left\|\textbf{h}_2\right\|^2 {<}\,x , \left\|\textbf{h}_{1}\right\|^2 {>}\,u \left\|\textbf{h}_{2}\right\|^2 \big| u\,{>}\,0 \right\rbrace \Pr\left\lbrace u_1{>}0,u_2{>}0 \right\rbrace \label{eqn:app:out2_2}\\ 
& \; + \Pr\left\lbrace\left\|\textbf{h}_2\right\|^2 {<}\,x , \left\|\textbf{h}_{1}\right\|^2 {>}\,u \left\|\textbf{h}_{2}\right\|^2 \big| u\,{<}\,0 \right\rbrace \Pr\left\lbrace u_1{>}0,u_2{<}0 \right\rbrace \label{eqn:app:out2_3}\\
& \; + \Pr\left\lbrace\left\|\textbf{h}_2\right\|^2 {<}\,x , \left\|\textbf{h}_{1}\right\|^2 {<}\,u \left\|\textbf{h}_{2}\right\|^2 \big| u\,{<}\,0 \right\rbrace \Pr\left\lbrace u_1{<}0,u_2{>}0 \right\rbrace \label{eqn:app:out2_4}\\
& \; + \Pr\left\lbrace\left\|\textbf{h}_2\right\|^2 {<}\,x , \left\|\textbf{h}_{1}\right\|^2 {<}\,u \left\|\textbf{h}_{2}\right\|^2 \big| u\,{>}\,0 \right\rbrace \Pr\left\lbrace u_1{<}0,u_2{<}0 \right\rbrace . \label{eqn:app:out2_5}
\end{align}
As before, the probability on the left in (\ref{eqn:app:out2_3}) turns into $\Pr\{\|\textbf{h}_2\|^2\,{<}\,x\}$, and (\ref{eqn:app:out2_4}) simply vanishes as $\|\textbf{h}_j\|^2\,{\geq}\,0$ is satisfied for any realization. Furthermore, combining (\ref{eqn:app:out2_2}) and (\ref{eqn:app:out2_5}) gives
\begin{align}
\Pr\left\lbrace\left\|\textbf{h}_2\right\|^2 {<}\,x , E_1 \big| u \right\rbrace &= \alpha_1 \Pr\left\lbrace\left\|\textbf{h}_2\right\|^2 {<}\,x \right\rbrace + \beta \Pr\left\lbrace\left\|\textbf{h}_2\right\|^2 {<}\,x , \left\|\textbf{h}_{1}\right\|^2 {>}\, u \left\|\textbf{h}_{2}\right\|^2 \big| u\,{>}\,0 \right\rbrace \,,\label{eqn:app:out2_6} 
\end{align}
and integrating (\ref{eqn:app:out2_6}) over the distribution of $\|\textbf{h}_1\|^2$ for $u\,{\geq}\,0$ produces
\begin{align}
\Pr\left\lbrace\left\|\textbf{h}_2\right\|^2 {<}\,x , E_1 \big| u \right\rbrace &= \alpha_1 F_\chi(x) + \beta \int_{0}^{\infty} \Pr\left\lbrace\left\|\textbf{h}_2\right\|^2 {<}\,x , \left\|\textbf{h}_{2}\right\|^2 {<}\,\sfrac[2pt]{h}{u} \right\rbrace f_\chi(h) \dd h \,,\nonumber \\
&= \alpha_1 F_\chi(x) + \beta \int_{ux}^{\infty} \Pr\left\lbrace\left\|\textbf{h}_2\right\|^2 {<}\,x \right\rbrace f_\chi(h) \dd h \nonumber\\
& \quad + \beta \int_{0}^{ux} \Pr\left\lbrace\left\|\textbf{h}_2\right\|^2 {<}\,\sfrac[2pt]{h}{u} \right\rbrace f_\chi(h) \dd h \,,\nonumber\\
&= \alpha_1 F_\chi(x) + \beta \left[ F_\chi(x)\left(1-F_\chi\left(ux\right)\right) + \int_{0}^{ux} \!\!\! F_\chi\left(\sfrac[2pt]{h}{u}\right) f_\chi(h) \dd h \right] . \label{eqn:app:out2_9}  
\end{align}
The second term in (\ref{eqn:app:out2_1}) can be evaluated by employing the methodology followed for the $1$st layer as follows
\begin{align}
\Pr\left\lbrace\left\|\textbf{h}_1\right\|^2 {<}\,x , E_2 \big| u \right\rbrace &= \Pr\left\lbrace\left\|\textbf{h}_1\right\|^2 {<}\,x ,  \left\|\textbf{h}_{1}\right\|^2 u_1\,{<} \left\|\textbf{h}_{2}\right\|^2 u_2 \right\rbrace \nonumber \\
&= \alpha_2 F_\chi(x) + \beta \left[ F_\chi(x)\left(1-F_\chi\left(\sfrac[2pt]{x}{u}\right)\right) + \int_{0}^{\frac{x}{u}} \! F_\chi(uh) f_\chi(h) \dd h \right] , \label{eqn:app:out2_11}
\end{align}
and the final expression for a given $u\,{\geq}\,0$ is obtained by combining (\ref{eqn:app:out2_9}) and (\ref{eqn:app:out2_11}) as follows
\begin{align}
F_2(x|u) &= (1-\beta) F_\chi(x) + 2 \beta \left[ F_\chi(x) - F_\chi(x)F_\chi(ux) + P\left(u,\sfrac[2pt]{x}{u}\right)\right]\,, \label{eqn:app:outage_15}
\end{align}
which agrees with (\ref{eqn:outage2_3}) after integration over the distribution of $u$.

Finally, we consider the integral function $P(u,a)$ in (\ref{eqn:pu_1}) assuming a chi-square distribution with $2N$ degrees of freedom and the associated cdf, $F_\chi(x)$, and the pdf, $f_\chi(x)$, given as
\begin{align}
F_\chi (x) &= 1-e^{-x/2}\sum\limits_{r=0}^{N-1} \frac{x^r}{2^{r}r!} \,, \label{eqn:chisquare_cdf}\\
f_\chi (x) &= \frac{x^{N-1}e^{-x/2}}{2^{N} (N-1)!} . \label{eqn:chisquare_pdf}
\end{align}
Employing (\ref{eqn:chisquare_cdf}) and (\ref{eqn:chisquare_pdf}) in (\ref{eqn:pu_1}) produces the following expression 
\begin{align}
P(u,a) &= F_\chi(a) - \frac{1}{2^{N}(N{-}1)!} \sum\limits_{r=0}^{N-1} \frac{u^r}{2^{r}r!} \int_{0}^{a} x^{N+r-1} e^{-\frac{1+u}{2}x} \dd x ,\label{eqn:app_pu_1}
\end{align}
where the definite integral in (\ref{eqn:app_pu_1}) can be evaluated with a help of the following identity~\cite{Gradshteyn}
\begin{align}
\int_{0}^{a} x^{m} e^{-cx} \dd x &= \frac{m!}{c^{m+1}} - e^{-ca} \sum\limits_{i=0}^{m} \frac{m!}{i!} \frac{a^i}{c^{m-i+1}} , \label{eqn:integral_identity}
\end{align}
by replacing $c\,{=}\,(1{+}u)/2$ and $m\,{=}\,N{+}r{-}1$. Employing (\ref{eqn:integral_identity}) in (\ref{eqn:app_pu_1}) yields (\ref{eqn:pu_2}) after some straightforward mathematical manipulations. 

At moderate to high SNR regime, we can safely approximate $P(u,a)$ around $u\,{=}\,1$ with a linear function of $1$st order polynomial given as 
\begin{align}\label{eqn:app_pu_2}
P(u,a) \simeq m_p (u-1) + b ,
\end{align}
which follows from the fact that $f_u(u)$ is non-zero only over the support set $\mathcal{U}_s\,{=}\,[1{-}\epsilon_1,1{+}\epsilon_2]$ with $\epsilon_i\,{\ll}\,1$, as is discussed in Appendix~\ref{app:fu_high_snr}. In (\ref{eqn:app_pu_2}), the slope $m_p$ can be found as
\begin{align*}
m_p &= \frac{\partial P(u,a)}{\partial u} \Big|_{u{=}1} = \sum\limits_{r{=}0}^{N{-}1} \binom{N{+}r{-}1}{r} \frac{N{-}r}{2^{N{+}r{+}1}} \, \left[ 1{-}\mathrm{e}^{{-}a} \sum\limits_{i{=}0}^{N{+}r{-}1} \frac{a^i}{i!} \left(1{-}\frac{i{-}a}{N{-}r}\right)^{i} \right] \,,
\end{align*}
after straightforward mathematical operations, and $b\,{=}\,P(1,a)$. 

\begin{footnotesize}
\bibliographystyle{IEEEtran}
\bibliography{IEEEabrv,paperbib}
\end{footnotesize}

\end{document}